\documentclass[%
 aip,jcp,
 sd,%
 amsmath,amssymb,
 reprint,%
]{revtex4-1}

\usepackage{graphicx}
\usepackage{dcolumn}
\usepackage{bm}

\usepackage{color}

\begin{document}

\title[WACSF - Weighted Atom-Centered Symmetry Functions as Descriptors in Machine Learning Potentials]{WACSF - Weighted Atom-Centered Symmetry Functions as Descriptors in Machine Learning Potentials}

\author{M. Gastegger}
\author{L. Schwiedrzik}
\author{M. Bittermann}
\author{F. Berzsenyi}
\author{P. Marquetand}
 \email{philipp.marquetand@univie.ac.at}
 \homepage{http://www.marquetand.net}
\affiliation{%
Institute of Theoretical Chemistry, Faculty of Chemistry, University of Vienna, W\"ahringer Str. 17, 1090 Vienna, Austria.
}%

\date{\today}

\begin{abstract}
We introduce weighted atom-centered symmetry functions (wACSFs) as descriptors of a chemical system's geometry for use in the prediction of chemical properties such as enthalpies or potential energies via machine learning. The wACSFs are based on conventional atom-centered symmetry functions (ACSFs) but overcome the undesirable scaling of the latter with increasing number of different elements in a chemical system. The performance of these two descriptors is compared using them as inputs in high-dimensional neural network potentials (HDNNPs), employing the molecular structures and associated enthalpies of the 133\,855 molecules containing up to five different elements reported in the QM9 database as reference data. A substantially smaller number of wACSFs than ACSFs is needed to obtain a comparable spatial resolution of the molecular structures. At the same time, this smaller set of wACSFs leads to significantly better generalization performance in the machine learning potential than the large set of conventional ACSFs. Furthermore, we show that the intrinsic parameters of the descriptors can in principle be optimized with a genetic algorithm in a highly automated manner. For the wACSFs employed here, we find however that using a simple empirical parametrization scheme is sufficient in order to obtain HDNNPs with high accuracy.
\end{abstract}

\maketitle

\section{\label{sec:introduction} Introduction}
The use of machine learning in quantum chemistry is currently trending.\cite{Behler2014JPCM,Pyzer-Knapp2015AFM,Behler2015IJQC,Gastegger2015JCTC,Behler2016JCP,Gomez-Bombarelli2016NM,Hase2016CS,Gastegger2016JCP,Yao2017JCP,Kobayashi2017PRM,Brockherde2017NC,Browning2017JPCL,Glielmo2017PRB,Janet2017JPCA,Behler2017ACIE,Ramakrishnan2017} One important domain is the description of the interatomic interactions by machine learning potentials,\cite{Behler2017ACIE,Ramakrishnan2017} which fit highly nonlinear analytical expressions to reference data obtained from electronic structure calculations. These machine learning potentials offer the advantage of high accuracy on par with ab initio methods and the ability to describe e.g. bond breaking and bond formation at a speed on par with classical force fields.\cite{Behler2016JCP} 

Not only potential energies but also other molecular properties, like atomization energies\cite{Rupp2012PRL} or dipole moments\cite{Gastegger2017CS}, can easily be modeled with such machine learning approaches. The applications are manifold, reaching from material design\cite{Pyzer-Knapp2015AFM} and scattering simulations\cite{Jiang2016IRPC} to the calculation of infrared spectra\cite{Gastegger2017CS}. Different machine learning variants for these targets can be found in the literature, with artificial neural networks\cite{Behler2011PCCP, Jiang2016IRPC} and kernel-ridge regression\cite{Rupp2012PRL,vonLilienfeld2015IJQC} being prominent examples. The common theme between electronic structure theory and these methods is that they take a molecular geometry as input and produce a molecular property as output. Since the electronic structure methods intrinsically contain the translational and rotational invariance of the electronic energies, it is advantageous to include this feature also in the machine learning methods.

The roto-translational invariance in the machine learning potentials is often achieved by a preprocessing step. The molecular geometries typically described in Cartesian coordinates are then transformed into other descriptors, i.e. other representations of the molecular structure. Not only roto-translational invariance can be achieved in this way but also other advantageous traits can be introduced, e.g., introducing cut-offs in order to restrict calculations to a certain spatial region of the investigated system and achieve a linear scaling behavior in the computational approach. There are infinite possibilities for defining such descriptors and the suitability of a descriptor may depend drastically on the machine learning model it is combined with.\cite{Faber2017JCTC} Accordingly, many different descriptors have been developed already but the search for better representations of molecular or condensed systems is a topic that is still gaining in attention. The existing models have been developed for neural networks or for kernel approaches (like kernel-ridge regression, support vector machines or Gaussian approximation potentials). In principle, all these descriptors could be used in the kernel approaches but not all of them are directly suited for neural networks, see below. 
In the following, we will discuss these developments in chronological order, without claiming to provide an exhaustive list. Note that in neighboring fields like quantitative structure-activity relationship (QSAR) research, a multitude of descriptors has been employed already for many years, comprehensively compiled e.g. in Ref.~\citenum{Todeschini2008}.

One of the obvious choices to obtain rotational and translational invariance in the descriptor is the use of internal coordinates, which have been adopted in early neural network implementations.\cite{Blank1995JCP,Brown1996JCP,Raff2005JCP}  In 2007, Behler and Parinello developed so-called high-dimensional neural network potentials (HDNNPs) and an important ingredient in these HDNNs are descriptors termed atom-centered symmetry functions (ACSFs), which are many-body functions based on radial and angular distribution functions.\cite{Behler2007PRL,Behler2011JCP} Rupp et al. used kernel-ridge regression to model atomization energies with a Coulomb matrix as descriptor, where the off-diagonal elements contain the Coulomb repulsion between two atoms and the diagonal elements correspond to a polynomial fit of atomic energies to a nuclear charge.\cite{Rupp2012PRL} Another notable descriptor is the smooth overlap of atomic positions (SOAP),\cite{Bartok2013PRB} which is an expansion of a local environment into spherical harmonics with atomic neighborhood densities and has been used in Gaussian approximation potentials,\cite{Bartok2015IJQC} but cannot directly be employed e.g. in neural networks, because it intrinsically contains a similarity measure between atomic neighborhoods. A descriptor employed by Kandathil et al. is termed atomic local frame (ALF), where a spherical polar coordinate-frame centered on an atom of interest is defined as the x-axis of the system, and the xy-plane, defined by selected, surrounding atoms \cite{Kandathil2013JCC}. Other descriptors include a bispectrum\cite{Bartok2013PRB}, metric fingerprints\cite{Sadeghi2013JCP}, partial radial distribution functions\cite{Schuett2014PRB}, internal representations based on force vectors\cite{Li2015PRL}, Fourier series of radial distribution functions\cite{vonLilienfeld2015IJQC}, bag of bonds\cite{Hansen2015JPCL}, graph fingerprints\cite{Duvenaud2015}, permutation invariant polynomials\cite{Jiang2016IRPC}, many-body expansions\cite{Huang2016JCP}, modified ACSFs\cite{Smith2017CS}, descriptors with constant complexity\cite{Artrith2017PRB}, spherical harmonics\cite{Jindal2017JCP}, simple elemental descriptors\cite{Seko2017PRB}, graph-based descriptors derived from SOAP\cite{Ferre2017JCP} and histograms of distances, angles or dihedrals\cite{Faber2017JCTC}.

These descriptors usually contain adjustable parameters like, e.g., the width of a Gaussian, which need to be predefined before the final representation is used as an input in the machine learning approach. Finding an optimal set of these parameters is a nontrivial problem and new approaches try to incorporate this task directly into the machine learning method.\cite{Schuett2017NC, Schuett2017arXiv, Lubbers2017arXiv} Some transformations are still done before a representation of the system's geometry is used in these methods but the general approach is promising.

In the present work, we adopt an alternative approach and first develop a new descriptor variant -- based on ACSFs but overcoming their undesirable scaling with the number of different chemical elements -- and afterwards optimize the intrinsic parameters with a genetic algorithm. The remainder of this work is structured as follows. In section~\ref{sec:theory}, the theoretical background and the newly introduced descriptor, termed weighted atom-centered symmetry function (wACSF), are presented. Computational details are given in section~\ref{sec:comput}. The wACSFs are applied in the prediction of enthalpies as comprised in the QM9 database\cite{Ramakrishnan2014SD} and the results are discussed in section~\ref{sec:results}. Specifically, the wACSFs are compared to ACSFs (section~\ref{sec:WACSF}), the influence of the parameters inherent to the descriptors is illustrated (section~\ref{sec:param}), and their optimization with a genetic algorithm is presented (section~\ref{sec:gen}), before summarizing the findings (section~\ref{sec:summary}).

\section{\label{sec:theory} Theory}

\subsection{High-Dimension Neural Network Potentials}

HDNNPs are a type of atomistic machine learning potentials developed by Behler and Parinello.\cite{Behler2007PRL}
In HDNNPs, a molecular property (e.g. potential energies, enthalpies) is obtained as the sum of individual atomic contributions.
These contributions depend on the local chemical environment of each atom and are modeled by artificial neural networks\cite{Bishop2006} (NNs).
Typically, a separate NN is used for every different chemical element present in the system under investigation.
An important feature of HDNNPs -- and atomic machine learning potentials in general -- is the way the local atomic environments are represented.
The three-dimensional structure of a molecule is encoded in the form of special atom-centered descriptors, which serve as inputs for the different elemental NNs.
These descriptors, as well as their influence on the overall quality of HDNNP models, are the focus of the present study and will be discussed in more detail in the next section.
For an in-depth description of HDNNPs, we refer to References~\citenum{Behler2014JPCM} and \citenum{Behler2015IJQC}.

\subsection{Atom-centered Symmetry Functions}

The primary type of descriptors used in HDNNPs are the aforementioned ACSFs.\cite{Behler2011JCP}
ACSFs model the local chemical environment of an atom $i$ via radial and angular distributions of the surrounding nuclei.

\begin{figure}[htpb]
\includegraphics[width=\columnwidth]{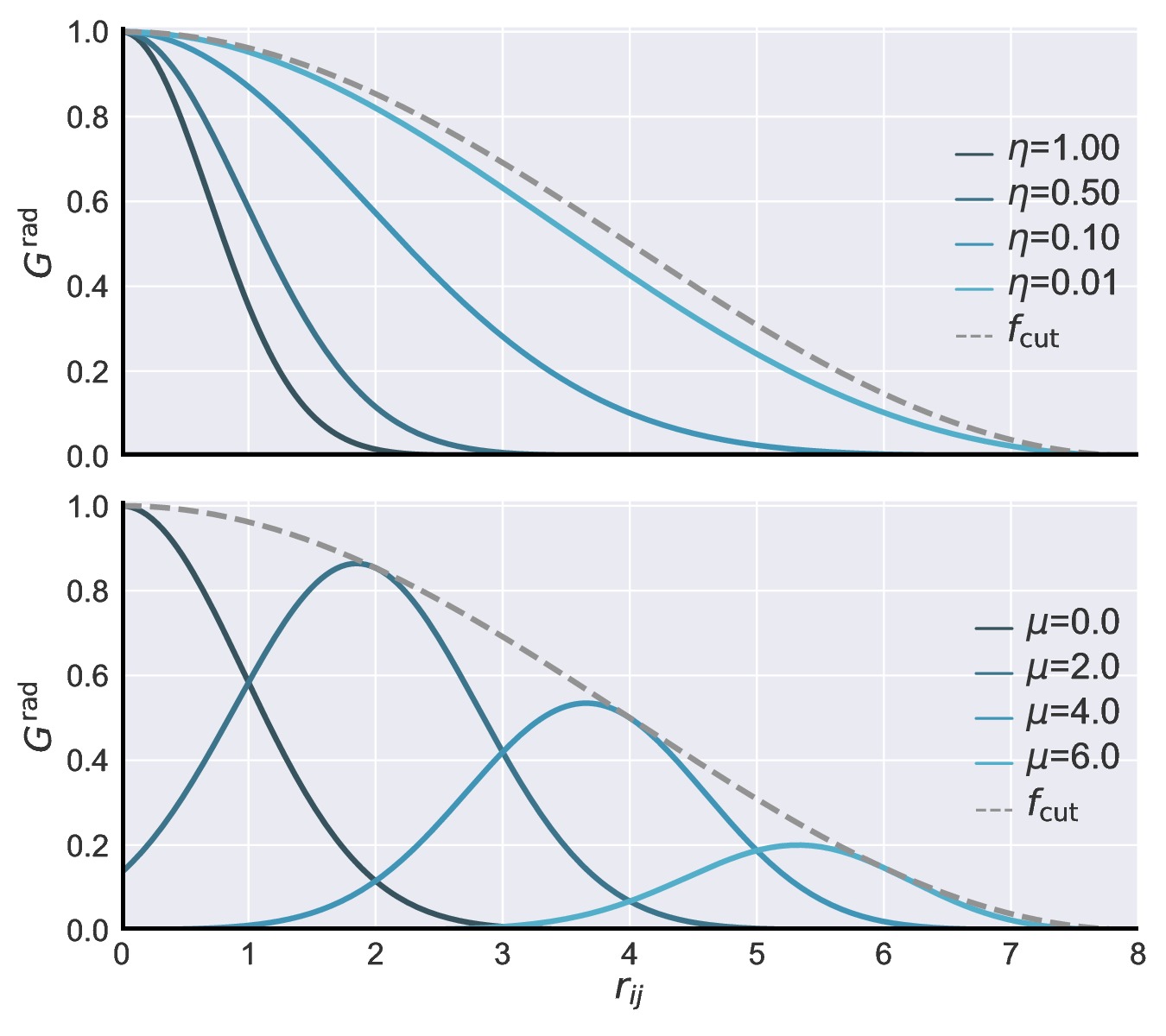}
\caption{\label{fig:rad} Examples for the influence of the parameters $\eta$ and $\mu$ on the overall shape of the Gaussian functions used in ACSFs and wACSFs. The upper panel shows Gaussians with $\mu=0$ using different widths $\eta$. In the lower panel, the width is kept constant and only the centers are shifted by varying $\mu$. The cutoff function $f_c$ controlling the overall region described by the function is depicted as a dashed, grey curve in both cases.}
\end{figure}

Radial ACSFs take the form
\begin{equation}
G^\textrm{rad}_i = \sum_{j \neq i}^N e^{-\eta (r_{ij}-\mu)^2 } f_\textrm{c}(r_{ij}), \label{eq:arad}
\end{equation}
where $N$ is the number of atoms and $r_{ij}$ is the distance between the atoms $i$ and $j$.
$\eta$ and $\mu$ are parameters modulating the width and position of the Gaussian function (Figure~\ref{fig:rad}).
A cutoff function $f_\textrm{c}$ ensures, that only the energetically relevant regions close to the central nucleus are encoded in the ACSF.
The most widely used cutoff function in the context of ACSFs is defined as:
\begin{equation}
f_{\textrm{c}}(R_{ij}) = 
\begin{cases}
\frac{1}{2} \left[ \cos \left( \frac{\pi r_{ij}}{r_{\mathrm{c}}} \right) + 1 \right],&  r_{ij} \leq r_{\textrm{c}}  \\
                                                                                   0,&  r_{ij} > r_{\textrm{c}},
\end{cases}
\end{equation}
where $r_c$ is the cutoff radius specifying the size of the region surrounding the central atom (see Figure~\ref{fig:rad}).

\begin{figure}[htpb]
\includegraphics[width=\columnwidth]{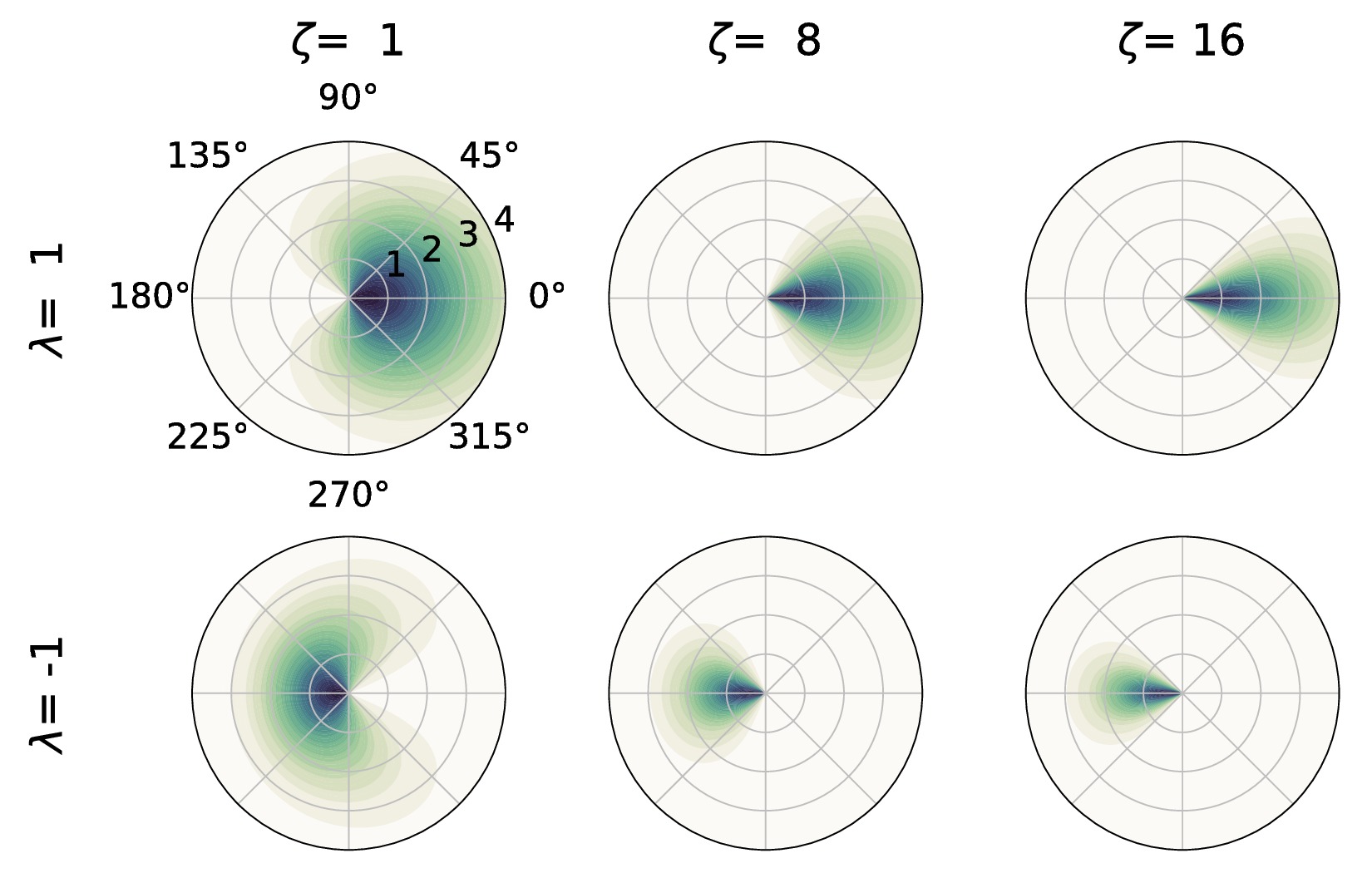}
\caption{\label{fig:ang} Polar plots depicting the influence of the parameters $\lambda$ and $\zeta$ on a single term of the sum in Equation~\ref{eq:aang}. All Gaussian functions are set to $\eta=0.01$. Changing the sign of the phase $\lambda$ moves the maximum of the angular density between 0$^\circ$ and 180$^\circ$. 
Increasing the parameter $\zeta$ focuses the function on a smaller range of angles close to the respective maxima. 
Although the top and bottom rows differ only in $\lambda$, a smaller spatial extent is observed for functions peaking at 180$^\circ$ compared to those with a maximum at 0$^\circ$.
This asymmetric behavior is due to the terms depending on $r_{jk}$ introduced in the angular descriptor.}
\end{figure}

Symmetry functions describing the angular environment take on a slightly more elaborate form:
\begin{eqnarray}
G^{\mathrm{ang}}_{i} &=& 2^{1-\zeta} \sum_{j \neq i}^N \sum_{k \neq i,j}^N \left( 1 + \lambda \cos\theta_{ijk} \right)^\zeta \nonumber \\
&& \times e^{-\eta (r_{ij}-\mu)^2} e^{-\eta (r_{ik}-\mu)^2} e^{-\eta (r_{jk}-\mu)^2} \nonumber \\ 
&& \times f_{ij} f_{ik} f_{jk}. \label{eq:aang}
\end{eqnarray} 
Here, $\theta_{ijk}$ is the angle spanned by the atoms $i$, $j$ and $k$. The term in brackets characterizes the distribution of angles. $\lambda$ is a parameter which takes the values $\lambda=\pm1$ and shifts the maximum of the angular term between $0^\circ$ and $180^\circ$, while $\zeta$ controls its width (Figure~\ref{fig:ang}).
$f_{ij}$, $f_{ik}$ and $f_{jk}$ are once again cutoff functions, where we have introduced the short hand notation $f_\textrm{c}(r_{ij})=f_{ij}$.
The introduction of terms based on $r_{jk}$ introduces asymmetric behavior into the angular functions, leading to a smaller spatial extent for angles close to 180$^\circ$ (see Figure~\ref{fig:ang}).

In order to describe the arrangement of different chemical elements surrounding the central atom, ACSFs are defined for pairs (radial) and triples (angular) of elements.
Only terms corresponding to these specific combinations of elements are counted in the summations in Equations~\ref{eq:arad} and \ref{eq:aang}.
For example, if a molecular system contains the elements H, C and O, the environment of a hydrogen atom would be described by a set of radial functions for the pairs H-H, H-C and H-O and 
angular functions for the triples H-H-H, H-H-C, H-H-O, H-C-C, H-C-O and H-O-O.
For each of these elemental combinations, several ACSFs are introduced respectively, varying in their parameters $\eta$, $\mu$, $\lambda$ and $\zeta$ (using e.g. a set of ACSFs with Gaussians of different widths $\eta$ to describe H-C distances).
This is done in order to provide a sufficient spatial resolution of all surrounding geometric features.
Finally, all these sets of radial and angular ACSFs are collected into a descriptor vector encoding the chemical environment of the central nucleus, based on which the elemental NNs make their predictions.

While ACSFs show an excellent performance for a wide range of systems, serious problems arise for molecules composed of several different chemical species.
Due to the way ACSFs are defined for combinations of elements, the number of functions necessary to describe a system depends directly on the number of elements $N_\textrm{elem}$ present. 
To account for every possible pair and triple, $N_\textrm{elem}$ radial and $N_\textrm{elem}(N_\textrm{elem}+1)$ angular symmetry functions are necessary (using two sets of angular functions with $\lambda= \pm 1$).
This number grows quickly, as the chemical composition of the molecules to be modeled increases in complexity.
While systems containing e.g. two elements can be described with eight unique combinations of ACSFs, 24 or 35 combinations are necessary when changing to four or five elements, respectively.
Although these numbers might seem small at a first glance, they only account for all possible combinations of elements.
In order to achieve a reasonable spatial resolution, several ACSFs are necessary for each of these combinations, further amplifying the undesirable scaling with $N_\textrm{elem}$. 
Continuing the above example, descriptor vectors of the lengths 40 (two elements), 120 (four elements) and 175 (five elements) would be obtained for every atom using e.g. sets of five ACSFs per pair and triple. 

The growing size of the descriptor vectors in turn leads to an increase in the computational cost associated not only with the training and evaluation of HDNNP models, but also with the transformation of the original Cartesian coordinates.
This behavior has serious implications for applications, where a large number of HDNNP evaluations are performed, e.g. high throughput screening or molecular dynamics simulations.
Due to these problems, standard ACSFs become less suited with an increasing number of different elements in chemical systems.

\subsection{Weighted ACSFs}

In order to overcome the limitations of conventional ACSFs discussed above, we propose a modification of this type of descriptor.
Instead of using separate functions to describe different combinations of elements, we instead account for the composition of the chemical environment in an implicit manner by introducing element-dependent weighting functions into Equations~\ref{eq:arad} and \ref{eq:aang}.
The resulting descriptors -- which we term weighted ACSFs (wACSFs) -- take the form
\begin{equation}
W^\textrm{rad}_i = \sum_{j \neq i}^N g(Z_j) e^{-\eta (r_{ij}-\mu)^2 } f_{ij} \label{eq:wrad}
\end{equation}
for radial and 
\begin{eqnarray}
W^\textrm{ang}_i &=& 2^{1-\zeta} \sum_{j \neq i}^N \sum_{k \neq i,j}^N h(Z_j,Z_k)  \left( 1 + \lambda \cos \theta_{ijk} \right)^\zeta \nonumber \\
&& \times e^{-\eta (r_{ij}-\mu)^2} e^{-\eta (r_{ik}-\mu)^2} e^{-\eta (r_{jk}-\mu)^2} \label{eq:wang} \nonumber \\ 
&& \times f_{ij} f_{ik} f_{jk}
\end{eqnarray}
for angular symmetry functions. $Z_j$ and $Z_k$ are the atomic numbers of the nuclei $j$ and $k$ respectively.
$g(Z_j)$ and $h(Z_j,Z_k)$ are weighting functions, which modify the contribution of each radial and angular term based on the chemical elements of the atoms involved.
While $g$ and $h$ can in principle use a wide variety of different definitions, we find that simply setting $g(Z_j)=Z_j$ and $h(Z_i,Z_j)=Z_i Z_j$ yields satisfying results without introducing additional parameters (possible alternatives for $h$ include e.g. $h(Z_j,Z_k) = \frac{Z_j Z_k}{Z_j + Z_k}$).

By directly incorporating information on the environment's elemental composition into the symmetry functions, the need for separate sets of functions for each combination of elements is eliminated. Hence, the number of wACSFs required to describe a system no longer depends on the number of different elements present, thus overcoming this inherent limitation of standard ACSFs.
A possible trade-off of wACSFs is the need for a finer variation in the parameters of the radial and angular functions in order to resolve the increased density of information contained in every individual wACSF (spatial and elemental). However, practical applications show that even when taking this effect into account, the wACSF descriptor vectors are still significantly shorter than their ACSF counterparts.

\subsection{Parametrization of Symmetry Functions}

Before HDNNPs can be applied to model a chemical system,  suitable sets of wACSF or ACSF parameters ($\eta$, $\mu$, $\lambda$ and $\zeta$) need to be determined first.
This task is typically performed in a trial and error fashion, involving knowledge on the system under investigation, as well as a certain degree of chemical intuition.
As a consequence, HDNNPs are hard to use out of the box without prior experience.
In the present work, we compare different empirical schemes for parametrizing wACSF and ACSF functions without the need for elaborate search procedures.

The choice of $\lambda$ and $\zeta$ for functions of the angular type is relatively straightforward. 
In general, it is beneficial to use two sets of angular functions with $\lambda=1$ and $\lambda=-1$ respectively, but identical parameters otherwise.
Since angular functions using these two values of $\lambda$ are complementary (see Figure~\ref{fig:ang}), all possible ranges of angles present in the environment can be resolved in this manner.
In a similar fashion, we use $\zeta=1$ for all wACSF and ACSF in the current work, as this provides a reasonable coverage of the angular space.
For larger $\zeta$, the descriptors focus increasingly on the regions close to $0^\circ$ and $180^\circ$, and the information on angles close to $90^\circ$ is lost.
However, in cases where the number of required angular functions far outweighs the radial ones, the inclusion of a few ACSFs or wACSFs with a higher value for $\zeta$ (e.g $\zeta=4$) can be of advantage.

By far the largest influence on descriptor performance is exerted by the parameters used in the radial parts of radial and angular symmetry functions.
These contributions are modeled by Gaussians, where the width $\eta$ and position $\mu$ of each Gaussian function modulates the spatial sensitivity of the corresponding descriptor.
Here, we investigate two alternative schemes for selecting appropriate sets of $\{\eta_i\}$ and $\{\mu_i\}$: The first scheme uses Gaussians of different widths centered at the origin (Figure~\ref{fig:rad}, upper panel), while the second one employs Gaussians of the same width, but shifted from the origin (Figure~\ref{fig:rad}, lower panel).

In both cases, we first choose the number of symmetry functions $N$.
In order to achieve a balanced coverage of the space until the cutoff radius $r_c$, we then introduce an auxiliary radial grid, based on which the function parameters are determined.
This auxiliary grid consists of $N$ equally spaced points $\{r_i\}$ ranging from $r_0$ to $r_N$. The distance $\Delta r$ between individual points in this grid is obtained as
\begin{equation}
\Delta r = \frac{r_N - r_0}{N - 1}.
\end{equation}

For the centered Gaussian functions, $r_0$ and $r_N$ are set to 1.0~\AA\ and $( r_\mathrm{c}-0.5 )$~\AA\ respectively.
After setting the shifts $\{\mu_i\}$ of all Gaussians to zero, the individual widths of each Gaussian are obtained as
\begin{equation}
\eta_i = \frac{1}{2 r_i^2}.
\end{equation}
Using this expression, the standard deviations of all Gaussian functions coincide with the points in the grid, thus fully covering the relevant radial range.

In the alternative scheme based on shifted functions, a lower limit of $r_0=0.5$~\AA\ is used.
Here, we assign the center of each individual Gaussian to one of the grid points, $\mu_i=r_i$.
Finally, the widths $\{\eta_i\}$ of the functions are chosen according to
\begin{equation}
\eta_i = \frac{1}{2 (\Delta r)^2}.
\end{equation}
In this way, the standard deviations of each Gaussian coincide with the centers of the two adjacent functions, thus achieving an extensive coverage of the region spanned by the grid, while maintaining a high radial resolution. 

The detailed analysis of the performance of the different parametrization strategies can be found in Section~\ref{sec:results}.

\subsection{Optimization of wACSF Descriptors with a Genetic Algorithm}

Recently, various approaches for the automated construction of descriptor vectors for atomistic NN potentials have been reported.
These are either based on heuristic search strategies, such as simulated annealing\cite{Jiang2013JCP} or incorporate the step of determining a set of optimal descriptors directly into the fitting procedure of the potential.\cite{Lubbers2017arXiv,Gomes2017arXiv,Schuett2017arXiv}

In the present study, we use a genetic algorithm (GA) to optimize wACSF descriptor vectors.
GAs are population based, metaheuristic optimization algorithms based on the principles of Darwinian evolution.\cite{darwin1859}
An initial generation of potential solutions is scored according to a fitness function. Based on this fitness score, a new generation is constructed using crossover and mutation operations.
This procedure is then continued in an iterative manner, until an optimal solution is found. For a more detailed discussion of GAs in general, we refer to Reference~\citenum{Goldberg1989}.

In GAs, the information pertaining to a particular solution is encoded in the form of a genome. In the case of wACSFs, the genome of a set of descriptors characterizing a system consists of entries containing the parameters $\eta$ and $\mu$ for the radial functions and $\eta$, $\lambda$ and $\zeta$ for the angular functions associated with each chemical element.
To cope with the special structure of the genome, small adaptations to the crossover and mutation procedure are introduced.
Operating on each symmetry function entry separately, the real-parameter simulated binary crossover scheme as described in Reference~\citenum{Deb96acombined} is used for the parameters $\mu$ and $\eta$, whereas the values of $\lambda$ and $\zeta$ are simply swapped between the parent genomes.
In a similar manner, we use the real mutation operation from Reference~\citenum{Deb96acombined} on the genome entries for $\eta$ and $\mu$.
Mutations on $\lambda$ and $\zeta$ are carried out by switching sign and incremental changes by $\pm1$ respectively. 
Crossover and mutation events occur stochastically according to predefined rates.

The parent genomes for the next generation are chosen via tournament selection according to the fitness of each individual genome present in the current generation.
Here, two alternative approaches to modeling the fitness function are pursued.
The first type of fitness function utilizes small HDNNP models to score a genome.
The evaluation of a genome in this manner proceeds along the following lines:
First, the GA generates a genome containing parameters for one set of symmetry functions. The reference structures in Cartesian coordinates are then transformed into the representation defined by these newly parametrized symmetry functions.
Based on the resulting representation, five individual HDNNPs are trained in sequence, using different strata of the reference set for training and validation (5-fold cross validation). The final fitness score of the genome is then obtained by averaging the validation mean absolute errors (MAEs) predicted by these models. 

In the second fitness function, the NNs in the HDNNPs are replaced by linear models obtained via linear ridge regression (LRR) instead.
This has the advantage, that the evaluation of an individual genome can be performed significantly faster, since the use of LRR is much cheaper than training even a small NN from a computational point of view.
However, since the optimized descriptors are to be used in conjunction with HDNNPs, the first approach is much closer to the original objective, potentially leading to improved performance.
Both fitness functions, as well as the performance of the GA in general, as well as the shape of the wACSF obtained in such a way are investigated in Section~\ref{sec:results}.

\section{\label{sec:comput} Computational Details}

Throughout this work, the QM9 database\cite{Ramakrishnan2014SD} was used as a reference data set for evaluating the different descriptor types, as well as the fitness functions guiding the GA.
The QM9 database contains the equilibrium structures and properties of 134\,855 small organic compounds composed of the elements H, C, N, O and F, which were computed at the B3LYP/6-31G(2df,p) level of theory.
This reference data was split into a training set consisting of 10\,000 randomly chosen structures and a test set containing the remaining molecules. 
All reported error measures were computed as the average of 5-fold cross validation\cite{Bishop2006}, using 8\,000 molecules for training, 2\,000 for validation and the remainder as test set.

All HDNNP models constructed in this work are based on NNs using two hidden layers with 10 and 50 nodes respectively, except when stated otherwise explicitly.
HDNNP training was performed with the element-decoupled Kalman filter\cite{Gastegger2015JCTC}, using an adaptive filter threshold of 0.9 and a time varying forgetting schedule of $\lambda_0=0.99$ and $\lambda_k=0.99$.
All models were trained for 100 epochs, except in the case of the HDNNP based GA fitness function, where only 20 epochs were used to reduce the time required for fitness evaluations.
The weights of the individual elemental NNs were initialized with the scheme reported by Glorot and Bengio.\cite{pmlr-v9-glorot10a}
In order to improve training performance, all descriptors were normalized to a mean of zero and a standard deviation of one.
Similarly, the enthalpies reported for the isolated atoms composing a molecule were subtracted from the corresponding total enthalpies of formation for every entry in the QM9 database.
In addition, the resulting enthalpy values were preprocessed further by normalizing to a standard deviation of one and subtracting a molecule specific contribution $N_i \bar{H}$, where $N_i$ is the number of atoms in molecule $i$ and $\bar{H}$ is the mean of all molecular enthalpies found in the training set divided by the associated numbers of atoms (see supporting information of Reference~\citenum{Schuett2017arXiv}).
For all symmetry functions a global cutoff of 8.0~\AA\ was used.
Descriptor transformation, as well as the training and construction of HDNNPs was carried out with an in-house python code utilizing the theano package.\cite{Rossum,TDT2016a}

The genetic algorithm is based on the GeneAS scheme described in Reference~\citenum{Deb96acombined}.
Using a population size of 100 individuals, 100 generations were evaluated during each GA run.
An exponent of $\eta=3.0$ was used in the real valued crossover and mutation routines, while the crossover and mutation rates were chosen as 0.5 and 0.08 respectively.
Individuals used for creating the next generation were selected with tournament selection using a tournament size of two.
In addition, an elitist scheme preserved the 20 best individuals found in every generation.
The fitness of individuals were evaluated based on the average MAEs computed via 5-fold crossvalidation on the training set.
With respect to the HDNNP based fitness score, elemental NNs using one hidden layer of 10 nodes were employed.
In case of the LRR fitness function, the fitness score was determined by decaying the ridge parameter from values of $10^{-2}$ to $10^{-7}$ in 50 steps and using the crossvalidation error of the best performing model obtained in this manner.

\section{\label{sec:results} Results and Discussion}

\subsection{\label{sec:WACSF} Comparison of ACSFs and wACSFs}

The main motivation for introducing wACSFs is to provide a balanced and concise description of systems composed of several different chemical species, thus overcoming one of the inherent limitations of conventional ACSFs.
In order to assess the performance of wACSFs compared to ACSFs for such systems, we use HDNNPs based on both descriptor types to model the enthalpies of formation for the molecules in the QM9 database.
The QM9 database is well suited for this task, as it contains 133\,855 organic compounds built from the five elements H, C, N, O and F,  giving rise to a wide range of different chemical motifs.

In the case of wACSFs, descriptor vectors of 32 symmetry functions were used to model the chemical environments of the individual atoms. Each descriptor is a combination of 26 shifted radial and 6 ($2\times3$) centered angular functions obtained with the parametrization schemes detailed in Section~\ref{sec:comput}.
For ACSF type functions, two different descriptor vectors were investigated.
The first one is a minimal descriptor, using on only one symmetry function for each combination of elements.
Since the QM9 database contains five different chemical elements, a total of 35 symmetry functions (5 shifted radial and 30 centered angular) is necessary
(see Section~\ref{sec:theory}), already exceeding the length of the wACSF vector.
However, in practical applications more than one ACSF is required per pair and triple of elements in order to resolve the geometries of the different chemical environments with sufficient accuracy.
Hence, we also introduce a second ACSF descriptor vector with an equivalent spatial resolution of symmetry functions as in the above wACSF descriptor vector.
In order to achieve this, 130 radial ($5\times26$) and 90 angular ($30\times6$) ACSFs are necessary, demonstrating the undesirable scaling of the overall length of ACSF type descriptor vectors with the number of chemical species.

\begin{figure}[htpb]
\includegraphics[width=\columnwidth]{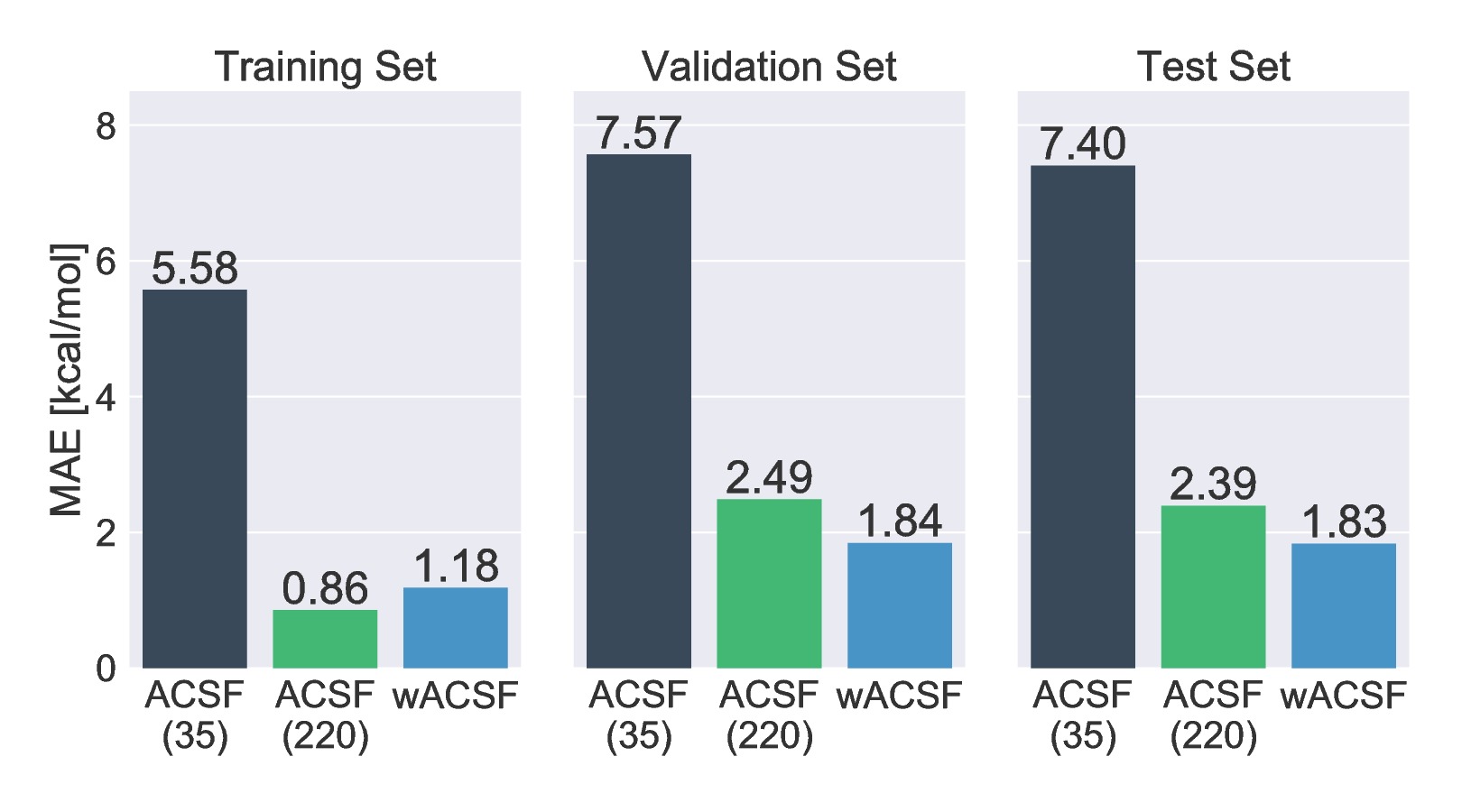}
\caption{\label{fig:compACSF} MAEs obtained for ACSF and wACSF type descriptors. In case of the wACSF based model, 32 symmetry functions are used to describe the chemical environment of each element (26 radial and 6 angular functions).
For the ACSFs, two different descriptor vectors are shown, one using a minimal set of 35 symmetry functions (5 radial and 30 angular), while the other one uses the same spatial resolution as the wACSF vector, leading to a total of 220 symmetry functions (130 radial and 90 angular). Note that better prediction errors may be achieved with larger descriptor vectors and larger NNs, but this was not the goal of this study.}
\end{figure}

Figure~\ref{fig:compACSF} shows the mean absolute errors (MAEs) for the QM9 database obtained with the wACSF descriptor (32 symmetry functions), as well as the minimal ACSF descriptor (35 symmetry functions) and the ACSF descriptor using an equivalent spatial resolution as for the wACSFs (220 symmetry functions). In all cases, the 5-fold cross validation procedure and reference data split described in Section~\ref{sec:comput} were used.
As can be seen, the wACSF based descriptor outperforms the minimal ACSF descriptor by more than a factor of three, although both use a similar number of symmetry functions.
This effect is due to the manner in which elemental information is incorporated into both descriptors.
Standard ACSFs introduce a separate set of symmetry functions for every relevant combination of elements (see discussion above). In the minimal ACSF descriptor, only one symmetry function per combination is used in order to obtain a descriptor vector of similar length as its wACSF counterpart. This restriction leads to an insufficient spatial resolution, resulting in the subpar performance of the minimal descriptor. In wACSF, different chemical species are accounted for in an implicit manner instead (see Section~\ref{sec:theory}).
Thus, the need for separate sets of functions is eliminated and wACSF can achieve a much higher spatial resolution than ACSF with the same number of symmetry functions, while still being able to successfully differentiate between elements.
Due to its higher spatial resolution, the second ACSF descriptor is able to achieve a performance comparable to the wACSFs, albeit at the cost of increasing the number of symmetry functions from 35 to 220.
However, even in this case, the wACSF based models exhibit better predictive power for unknown samples, as can be seen in the MAEs associated with the validation and test sets, which are more than 0.5 kcal/mol lower than those of the ACSF descriptor using 220 symmetry functions (Figure~\ref{fig:compACSF}).
This finding demonstrates a second important feature of wACSF type functions.
In molecular systems, chemical motifs occur with different frequencies, it is e.g. much more likely to encounter bonds involving oxygen than fluorine in the QM9 database.
Since ACSFs treat different combinations of elements separately, models based on this descriptor type can only learn the contributions of rare motifs (e.g. an C-F bond) based on the few examples present in the training set. All information about patterns, which are structurally similar but involve different species (e.g. C-O bonds with a similar length as C-F bonds), is effectively wasted. 
In wACSF on the other hand, spatial functions are shared between different elements.
Hence, even if an element is only encountered rarely, wACSF based models are able to utilize the information about structurally similar motifs during training.
This feature is conceptually similar to the sharing of weights in modern neural network architectures and leads to the improved generalization behavior observed above.

In general, we find that for chemically diverse systems the newly introduced wACSF type symmetry functions offer superior performance compared to conventional ACSFs, not only with respect to their overall predictive accuracy, but also the total number of functions needed to describe a system.
The latter feature is especially beneficial for practical applications such a molecular dynamics simulations or high-throughput screening, as shorter descriptor vectors lead to a significant reduction in computational cost.

\subsection{\label{sec:param}Comparison of Parametrization Strategies}

After having demonstrated the overall advantages of wACSF, we move on to discuss different parametrization schemes for this kind of descriptor.
As was detailed in Section~\ref{sec:theory}, the present work differentiates between two main parametrization strategies for the spatial part of symmetry functions:
1) Gaussian functions which share the same center but use different widths (referred to as ``centered'') and 2) Gaussian functions of the same width but shifted along the spatial dimension (called ``shifted'').

\begin{figure}[htpb]
\includegraphics[width=\columnwidth]{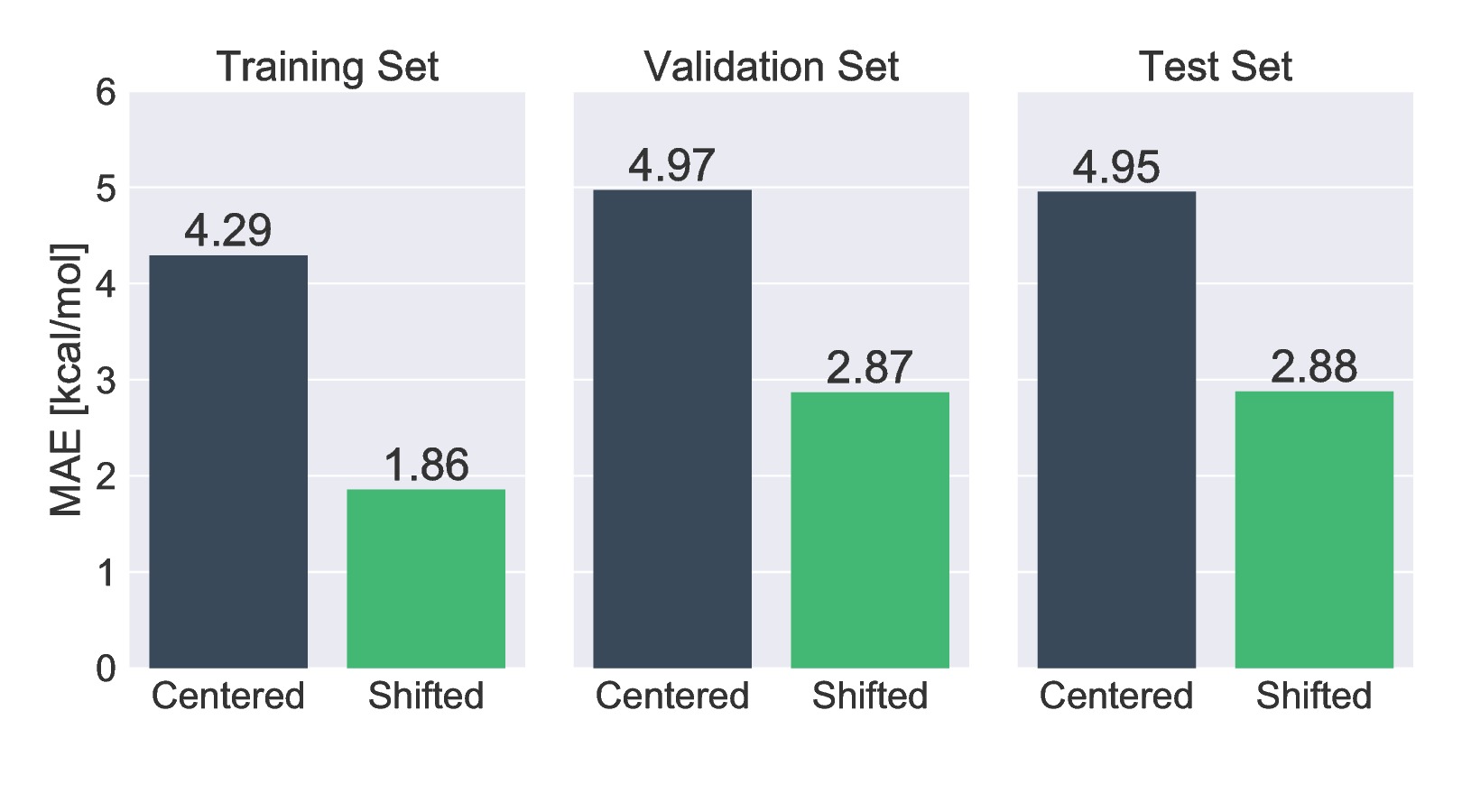}
\caption{\label{fig:comprad} Comparison of the performance of radial wACSFs employing centered and shifted Gaussian functions. In all cases, the descriptors based on the shifted parametrization outperform their counterparts, due to the better radial resolution offered by these type of functions.}
\end{figure}

Figure~\ref{fig:comprad} shows the performance of radial wACSF descriptors using 32 radial functions based on both schemes for the QM9 database.
As can be seen, using a set of shifted Gaussian functions offers a much higher accuracy than centered functions.
In the latter case, the individual Gaussians form concentric spheres around the central atom, resulting in large regions of overlap.
Hence, most symmetry functions contribute a signal over a wide range of radial positions, making the contributions of individual atomic positions within the environment less distinct.
Shifted Gaussians, however, are localized to specific shells around the central atom and therefore only provide signals at very distinct radial positions.
This improved radial resolution is the reason for the higher accuracy observed for radial wACSF using shifted functions.

\begin{figure}[htpb]
\includegraphics[width=\columnwidth]{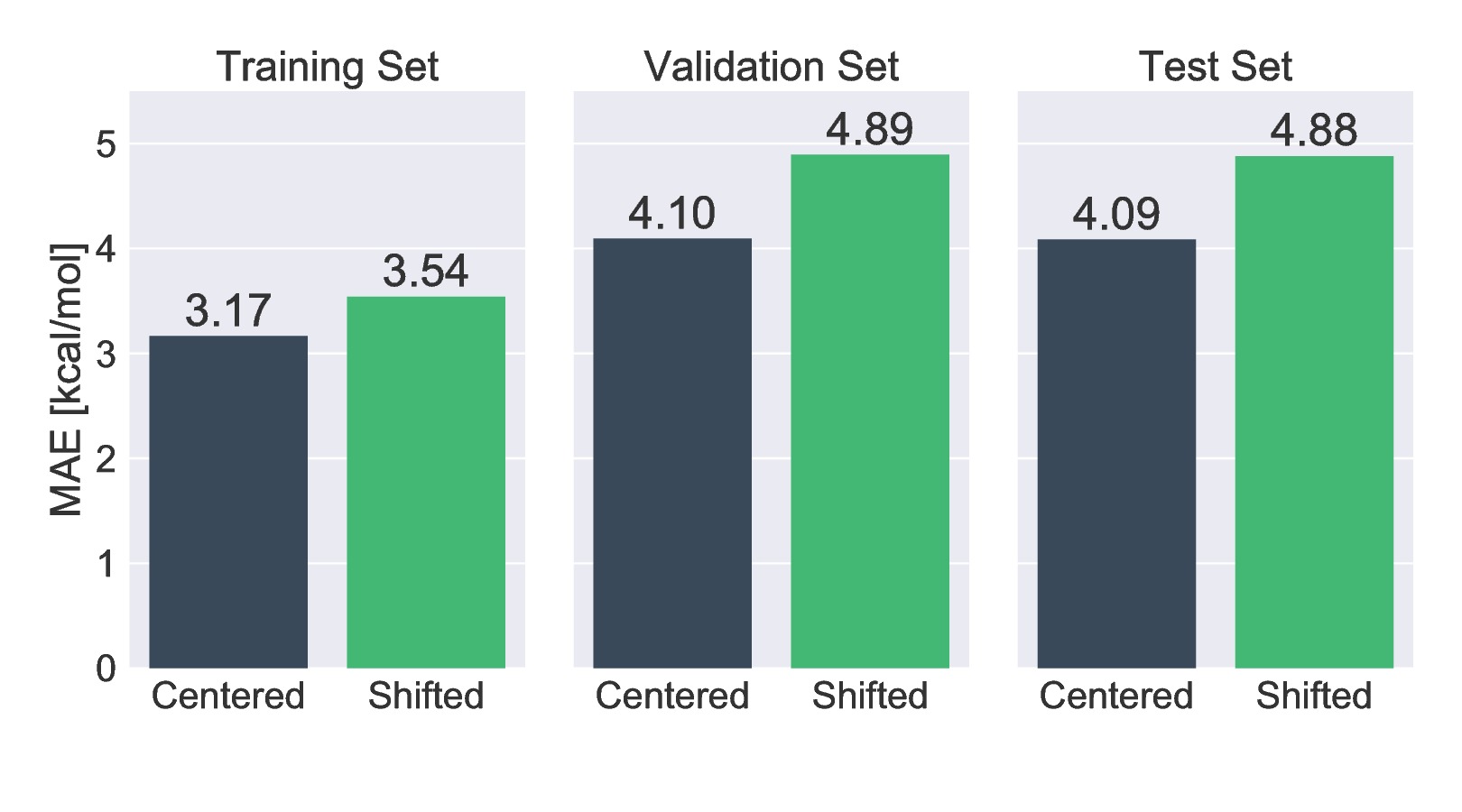}
\caption{\label{fig:compang} Difference between angular wACSFs using shifted and centered Gaussian spatial components. Unlike in the radial functions, here using the centered scheme offers a small, but distinct advantage.}
\end{figure}

For angular wACSFs, the situation is different. Figure~\ref{fig:compang} once again shows the performance of descriptors using both parametrization strategies for QM9, but this time using 32 angular instead of radial wACSFs.
Unlike in the case of radial functions, the use of concentric Gaussians is advantageous in combination with angular wACSF descriptors.
This behavior is due to the way the different types of wACSF are formulated. While radial functions only use sums of Gaussian functions, their angular counterparts involve the product of three Gaussians (see Section~\ref{sec:theory}).
In the case of the highly localized Gaussians used in the shifted parametrization scheme, this product vanishes easily, due to the small spatial extent and consequently negligible overlaps of each Gaussian.
As a consequence, information about the spatial structure of the central atoms environment is discarded in many cases, leading to the suboptimal performance of angular wACSFs using this strategy.
Here, the large regions of overlap between the concentric Gaussian functions prove to be advantageous. Due to the large spatial extent of each Gaussian, the aforementioned product only vanishes completely in the most extreme cases and more spatial information of the environment can be encoded. The result is the improved accuracy for the centered functions compared to the shifted functions, as can be seen in Figure~\ref{fig:compang}.
While we focused our analysis on wACSFs type functions, the same observations as above also hold in the case of ACSFs.

\subsection{Ratio of Radial to Angular Symmetry Functions}

Conventional ACSF descriptors use a combination of radial and angular symmetry functions in order to describe the chemical environments present in a molecule.
Here, we study in how far wACSF type descriptors profit from the same strategy and whether an optimal ratio of radial to angular functions exists.

\begin{figure}[htpb]
\includegraphics[width=\columnwidth]{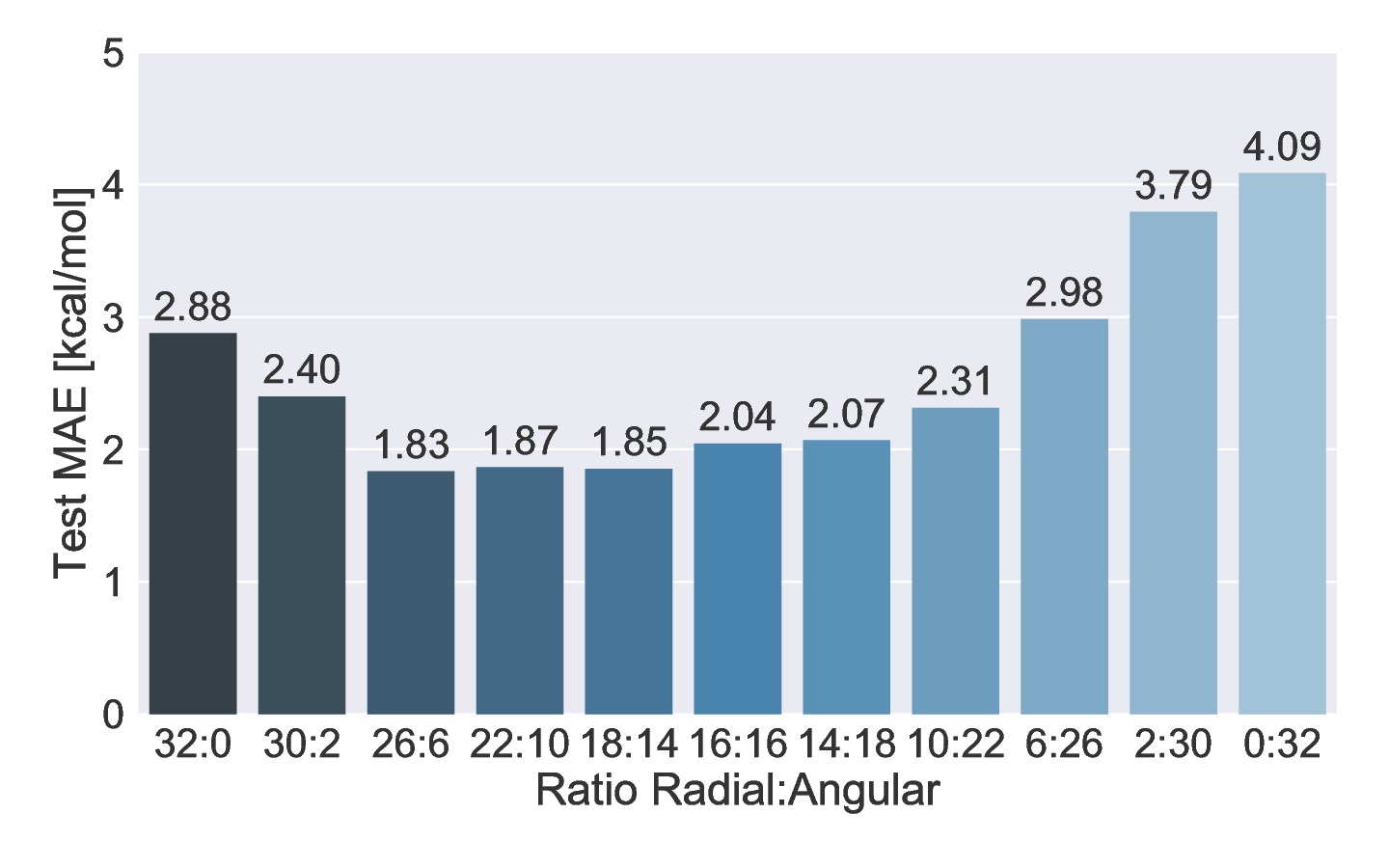}
\caption{\label{fig:ratio} Test set MAEs obtained for descriptor vectors using different ratios of radial to angular functions. In all cases, the total number of symmetry functions is kept constant at 32. Since blocks of $\lambda=\pm1$ are used, the number of angular functions can only vary in steps of two. For reasons of clarity, not all possible combinations are shown (a figure considering all ratios can be found in the supporting information).}
\end{figure}

\begin{figure}[htpb]
\includegraphics[width=0.9\columnwidth]{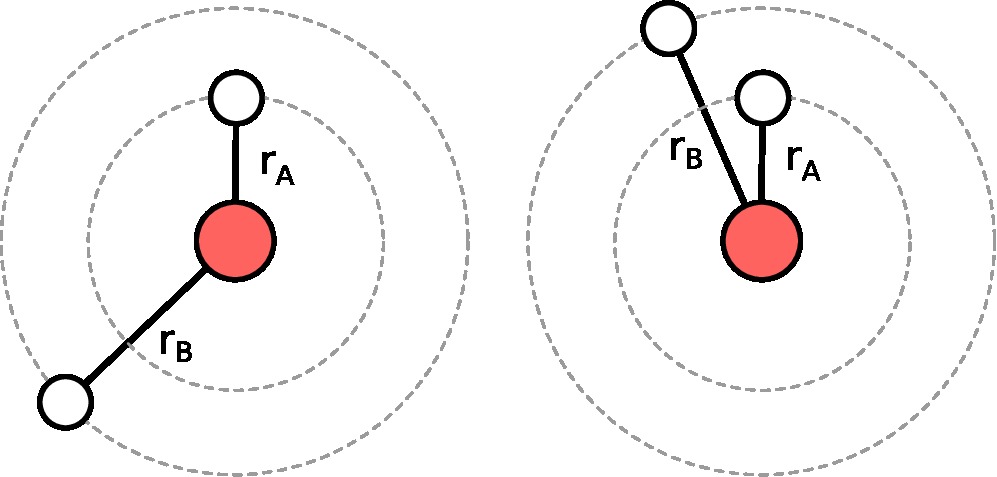}
\caption{\label{fig:prob} Since radial symmetry functions depend only on the distances to the neighboring atoms ($r_A$ and $r_B$), they are unable to distinguish between the two environments depicted above, although both are chemically distinct.}
\end{figure}

To this end, we apply HDNNP models based on descriptor vectors using a different number of shifted radial and centered angular symmetry functions to the QM9 dataset and monitor the respective MAEs predicted for the test set (see Figure~\ref{fig:ratio}).
Looking at the extreme cases first, we find that the use of descriptor vectors consisting of only angular functions (0:32) results in the worst performance of all studied cases.
The cause of this behavior is the fact that these descriptors use centered Gaussian functions and hence sacrifice radial resolution in favor of the ability to describe the angular distributions of atoms in the environment (see discussion above).
Radial wACSFs (32:0) on the other hand, provide a much more refined description of the spatial environment and appear to be able to capture the majority of the geometric features present in the QM9 database, as is attested by their relatively low MAEs compared to the purely angular descriptors. 
However, radial wACSFs on their own suffer from an inherent limitation, as they are unable to describe relative positions between atoms surrounding the central atom.
An example for this behavior is given in Figure~\ref{fig:prob}, which shows two systems where atoms are positioned at the same distance from the center, albeit at different angles.
Since the sum in the definition of radial wACSFs (see Equation~\ref{eq:wrad}) only depends on the relative distances between the respective central atom and its neighbors,
exactly the same descriptor values are obtained in the two cases.
Hence, although both systems are completely distinct from a chemical point of view, radial wACSFs are unable to distinguish between them.

In order to differentiate between such environments, the information provided by angular symmetry functions is crucial.
This effect can be seen in Figure~\ref{fig:ratio}, where the inclusion of only one pair of angular functions in the descriptor vector already leads to a noticeable gain in accuracy.
Increasing the ratio of angular to radial wACSF even further, the performance continues to improve, yielding the best set of descriptors found in this work (26:6).
Past a certain point (18:14), the quality of the HDNNPs models decreases again, as the loss in radial resolution begins to outweigh the angular information gained.

All in all, the dependence of wACSF performance on the ratio of symmetry functions exhibits a very systematic behavior.
We expect, that this trend also holds for chemical problems other than QM9, albeit with the minimum situated at a different ratio.
This is due to the composition of QM9, as this database only holds equilibrium structures. For other applications, e.g. molecular dynamics, a larger amount of angular functions will most likely be beneficial in order to resolve the slight geometric variations incurred due to a molecules change over time.
Finally, it should also be noted that the performance of purely angular wACSF can be improved by varying the $\zeta$ parameter (see Equation~\ref{eq:wang}). However, in the present work we strive to find parametrization schemes which can be used out of the box, which was why a detailed analysis of this parameter was not conducted.

\subsection{\label{sec:gen} Genetic Algorithm Optimization of wACSF}

Until now, this study has only focused on wACSF descriptors obtained with simple, problem independent parametrization strategies.
However, in practical applications of HDNNPs, a set of descriptors is typically tailored to a specific chemical problem via a tedious trial and error procedure.
Here, we investigate whether the search for optimal descriptors can be automatized with the help of a genetic algorithm and inhowfar the performance of wACSF vectors selected in such a manner differs from one of the unoptimized descriptors.

\begin{figure}[htpb]
\includegraphics[width=\columnwidth]{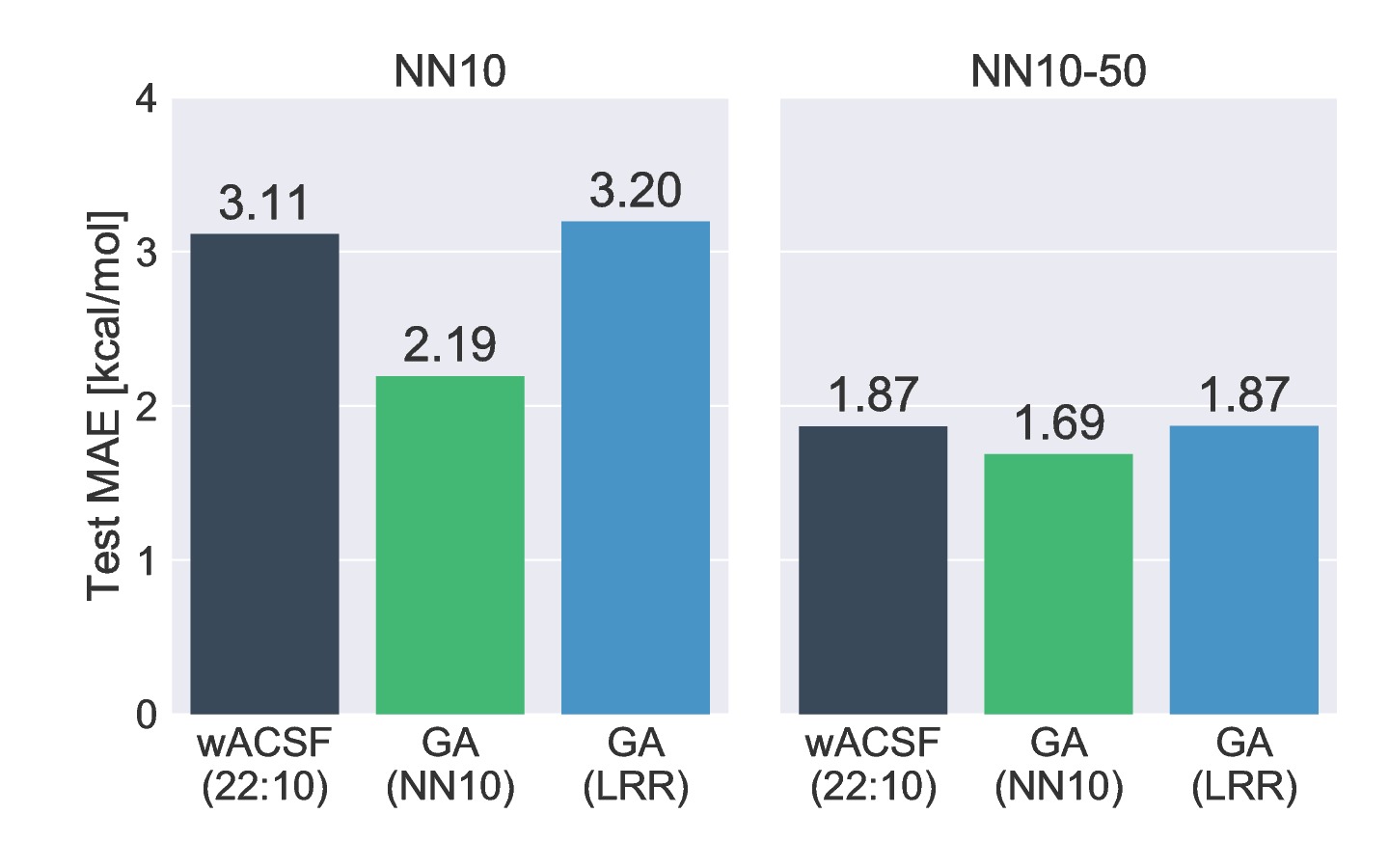}
\caption{\label{fig:GA} Performance of the descriptors optimized with the GA using fitness functions based on linear ridge regression (GA-LRR) and HDNNPs (GA-NN10). Shown are the test set MAEs obtained for HDNNP models based on elemental networks using a single hidden layer of 10 nodes (NN10), as well as two-hidden-layer architectures using 10 and 50 nodes (NN10-50).
The performance of the original descriptor (wACSF) is shown in black.}
\end{figure}

To this end, we use the genetic algorithm described briefly in Section~\ref{sec:theory} to optimize a set of wACSFs to be used in HDNNP models of the QM9 enthalpies.
A wACSF descriptor based on 22 radial and 10 angular functions (see Figure~\ref{fig:ratio}) is used as a starting point and all parameters other than the cutoff radius are optimized with the genetic algorithm.
Two alternative choices of fitness functions are studied, one using a model obtained via linear ridge regression (LRR) and the other a fit based on HDNNPs formed from small single layer NNs using 10 nodes (see Section~\ref{sec:theory}).
The final performance of the optimized descriptor vectors was evaluated based on the test MAEs for QM9 obtained for the respective HDNNP models via 5-fold cross validation.
Two different HDNNP architectures were used in this evaluation procedure, one based on two-hidden-layer NNs using 10 and 50 nodes (NN10-50), which was also employed previously, and a second one using smaller single layer networks with only 10 hidden nodes (NN10). Due to the high associated computational cost, no GA optimization were carried out using a fitness based on the 10-50 HDNNPs directly. All associated results are depicted in Figure~\ref{fig:GA}.

Focusing first on the MAEs achieved by the smaller HDNNPs (NN10), we find that the genetic algorithm guided by the HDNNP based fitness function does indeed lead to a substantial improvement in accuracy, lowering the test MAE from 3.11 kcal/mol to 2.19 kcal/mol. These results are comparable to those of the previously studied models, with the MAE only being 0.32~kcal/mol higher than the one obtained for the best performing model (26:6), despite the reduction in NN size.
This finding demonstrates the importance of choosing an appropriate descriptor vector in combination with relatively small NN architectures.
Although larger HDNNPs (NN10-50) trained on the same descriptor still profit from the optimization procedure, the overall gain in performance is significantly smaller than compared to the NN10 models (improvement of only 0.18~kcal/mol from 1.87~kcal/mol to 1.69 kcal/mol).
This behavior can be related to the ability of the larger ML model to perform an internal transformation of the descriptor vector.
Small models only possess a limited fitting capacity and use most of their internal resources to model the relationship between the descriptor vector and the target properties, in this case the QM9 enthalpies.
However, if the number of free parameters in the model (e.g. network weights) grows larger, the resulting additional capacity can be used to transform a suboptimal descriptor vector to a more suitable representation.
As a consequence, the overall accuracy of HDNNPs (and ML models in general) of reasonable size is much less dependent on a specific parametrization of the descriptor, as long as it encodes sufficient information on structural and elemental patterns present in the reference data set.

Pertaining to the choice of fitness function, we find that using the HDNNP based fitness measure is generally more reliable than the LRR alternative.
Although the GA improves the overall fitness measure yielded by the LRR model from 5.67 kcal/mol to 5.18 kcal/mol during optimization, no improvement in accuracy is observed when the resulting descriptor vectors are used in conjunction with HDNNPs. In the case of NN10, the LRR based fitness measure even leads to a deterioration of the model quality.
While this kind of fitness function can show similar performance to the HDNNP alternative in some cases (see supporting information), its inconsistent behavior makes it less suitable as a fitness measure than the latter approach.

\begin{figure}[htpb]
\includegraphics[width=\columnwidth]{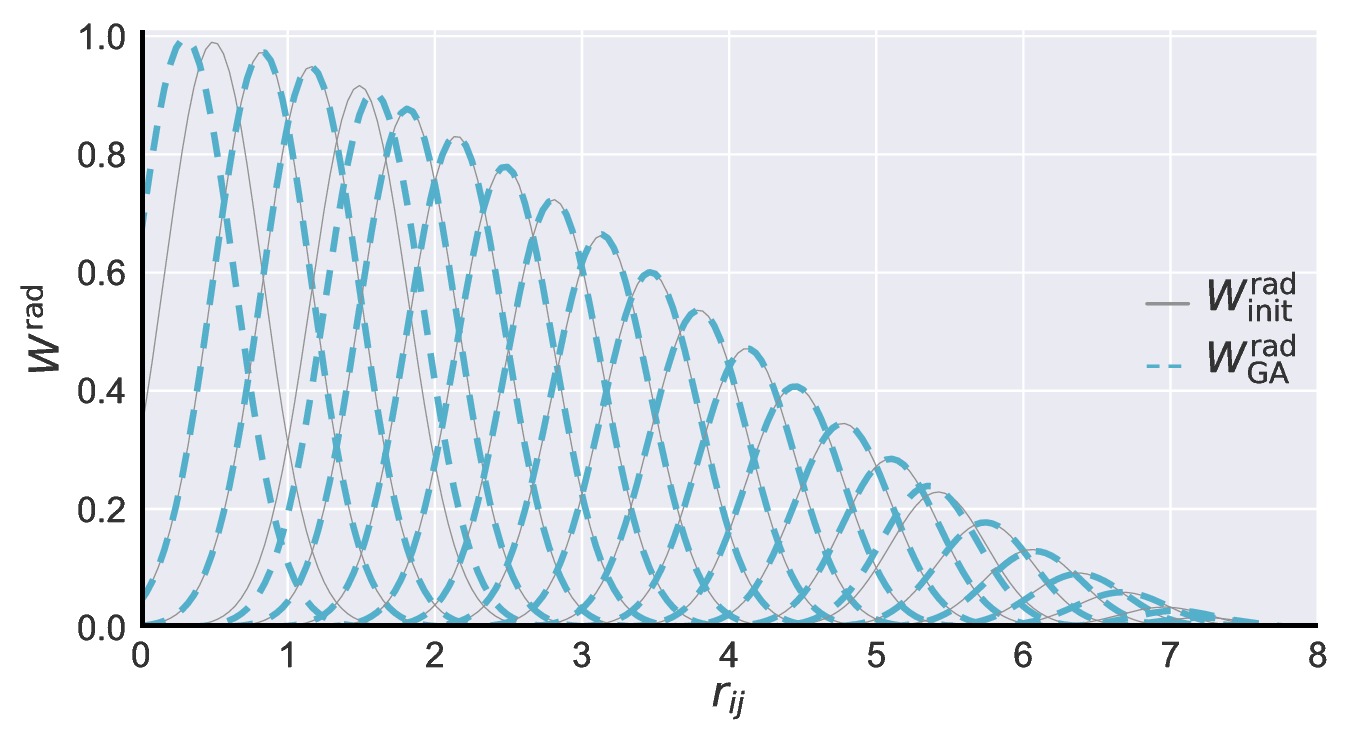}
\caption{\label{fig:diffrad} Radial components of wACSF vector (22:10) before ($W_\mathrm{init}^\mathrm{rad}$) and after ($W_\mathrm{GA}^\mathrm{rad}$) optimization with the GA.}
\end{figure}

\begin{figure*}[htpb]
\includegraphics[width=\textwidth]{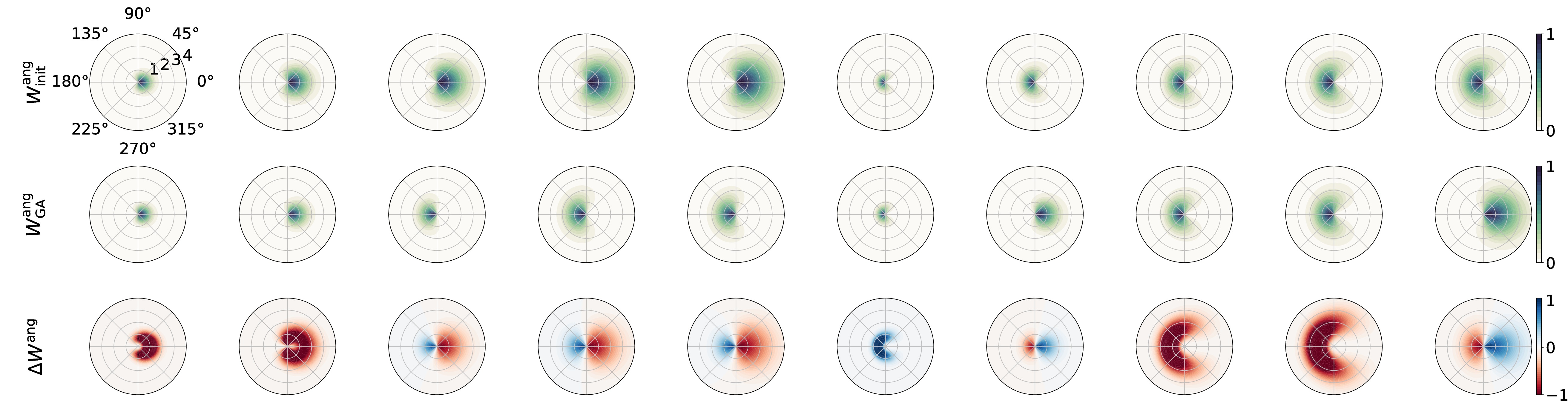}
\caption{\label{fig:diffang} Polar representations of the angular functions used in the wACSF descriptor (22:10) before ($W_\mathrm{init}^\mathrm{ang}$) and after ($W_\mathrm{GA}^\mathrm{ang}$) optimization with the GA. The lowest panel depicts the differences introduced by the GA ($\Delta W^\mathrm{ang} = W_\mathrm{GA}^\mathrm{ang} - W_\mathrm{init}^\mathrm{ang}$). The differences have been normalized to a range from -1 to 1 for visualization purposes.}
\end{figure*}

The differences between radial and angular components of the original ($W_\mathrm{init}$) and optimized wACSF vectors ($W_\mathrm{GA}$) are shown in Figures~\ref{fig:diffrad} and \ref{fig:diffang}, using the descriptors of the chemical environment of carbon as an example.
As can be seen in Figure~\ref{fig:diffrad}, the GA induces only minor changes into the radial symmetry functions, a trend which is also observed for the other elements (see supporting information).
This behavior indicates, that the radial wACSFs yielded by the parametrization scheme described above are already close to an optimum, at least in case of the QM9 database.
The most marked differences between $W_\mathrm{init}^\mathrm{rad}$ and optimized wACSF vectors $W_\mathrm{GA}^\mathrm{rad}$ can be found in the shifts of the Gaussian functions $\mu$.
Here, especially the change associated with the function closest to the center is pronounced, which moves to even shorter distances. This shift can be interpreted as an attempt by the GA to provide additional resolution for the spatial regions of C-H bonds.
Due to the different substitution patterns present in QM9, these bonds can show very small variations centered around 1.09~\AA. 
Hence, moving the outermost part of the Gaussian which exhibits the highest curvature towards the relevant bond lengths allows for a finer differentiation between these chemical species.

Pertaining to the angular wACSFs, much larger changes are found (see Figure~\ref{fig:diffang}).
The widths $\eta$ of the radial Gaussian components are generally shifted to much shorter distances compared to the unoptimized functions as is indicated in the difference plot $\Delta W^\mathrm{ang}$.
Another interesting observation is related to the phase of the angular term $\lambda$. Although $\lambda$ is switched by the genetic algorithm in several cases, a ratio close to 1:1 (4:6 for carbon) is still conserved.
However, the overall symmetry between the function blocks associated with $\lambda=\pm1$ generated by the parametrization scheme is now broken, as both sets vary in their parameters.
Finally, only small changes are observed for the angular exponent $\zeta$ (typically from $\zeta=1$ to $\zeta=2$), resulting only in marginal differences in the width of the angular distributions.
This finding supports our previously established convention of choosing $\zeta=1$ for all functions in the empirical parametrization scheme.

\section{\label{sec:summary} Summary}

In the present work, we introduce a new descriptor for atomistic neural network potentials (NNPs) of the Behler--Parrinello type.\cite{Behler2007PRL}
This descriptor -- termed weighted atom centered symmetry functions (wACSFs) -- is an adaptation of conventional ACSFs designed with the goal to overcome the shortcomings of the latter when faced with systems containing a moderate to large number of different chemical elements.

Using the molecules and associated enthalpies of the 133\,855 molecules reported in the QM9 database\cite{Ramakrishnan2014SD} as a reference, we find that the adapted descriptor exhibits excellent performance compared to conventional ACSFs. When comparing ACSF and wACSF descriptor vectors of similar length, a reduction from 7.40~kcal/mol (ACSFs) to 1.83~kcal/mol (wACSFs) is observed for the mean absolute prediction errors (MAEs). Note that better prediction errors may in principle be achieved with larger descriptor vectors and larger NNPs but this was not the goal of this study.
The wACSF type functions show a significantly better generalization performance than ACSFs with a comparable spatial resolution, while at the same time requiring a much smaller number of symmetry functions for a description of the reference systems (32 vs. 220 functions in the chosen example).
In addition to introducing wACSF type functions, we find that using simple empirical parametrization schemes potentials of high accuracy can be obtained out of the box, without the need for a tedious search procedure. 

Finally, the use of a genetic algorithm for optimizing wACSF descriptor vectors in a highly automated manner is analyzed.
Here, we find that relatively small neural network potentials profit greatly from this procedure (reducing the overall MAE from 3.11 to 2.19~kcal/mol for our 10-NNP).
However, when applied to models exhibiting a larger number of free parameters, the gain in accuracy is negligible compared to the additional computational effort (from 1.87 to 1.69~kcal/mol for a 10-50-NNP).
This finding supports the observation that sufficiently large NNs are able to internally transform suboptimal descriptor vectors to more suitable representations, a fact which is exploited in models like those reported in References~\citenum{Schuett2017arXiv}, \citenum{Lubbers2017arXiv} and \citenum{Gomes2017arXiv}.

In summary, wACSFs constitute a robust and economic choice of descriptors for atomistic potentials without the need for much additional optimization, especially when combined with the suggested parametrization schemes.
Although the QM9 represents a relatively incomplete snapshot of the vast chemical compound space, the observed performance of wACSFs suggests that the above trends will also hold for future studies aimed at non-equilibrium structures and molecular forces.

\section*{Supplementary Material}

See supplementary material for a full scan of the ratio of radial to angular functions, the genetic optimization of an additional set of wACSFs, the comparison of the genetically optimized wACSFs vs. the unoptimized ones for the elements H, N, O, F, and a listing of the parameters used in the best performing wACSFs in the present work.

\section*{Acknowledgement}
We thank Michael Seifner for preliminary work on wACSFs.
Allocation of computer time at the Vienna Scientific Cluster (VSC) is gratefully acknowledged.
MG is thanks the Monatshefte f\"ur Chemie for financial support via the MoChem scholarship.

%

\clearpage

\renewcommand{\thefigure}{S\arabic{figure}}
\renewcommand{\thetable}{S\arabic{table}}
\setcounter{figure}{0} 
\setcounter{table}{0} 

\section*{Supporting Information}
\subsection*{Full Scan of Radial to Angular Ratio}

\begin{figure}[htpb]
\centering
  \includegraphics[width=\columnwidth]{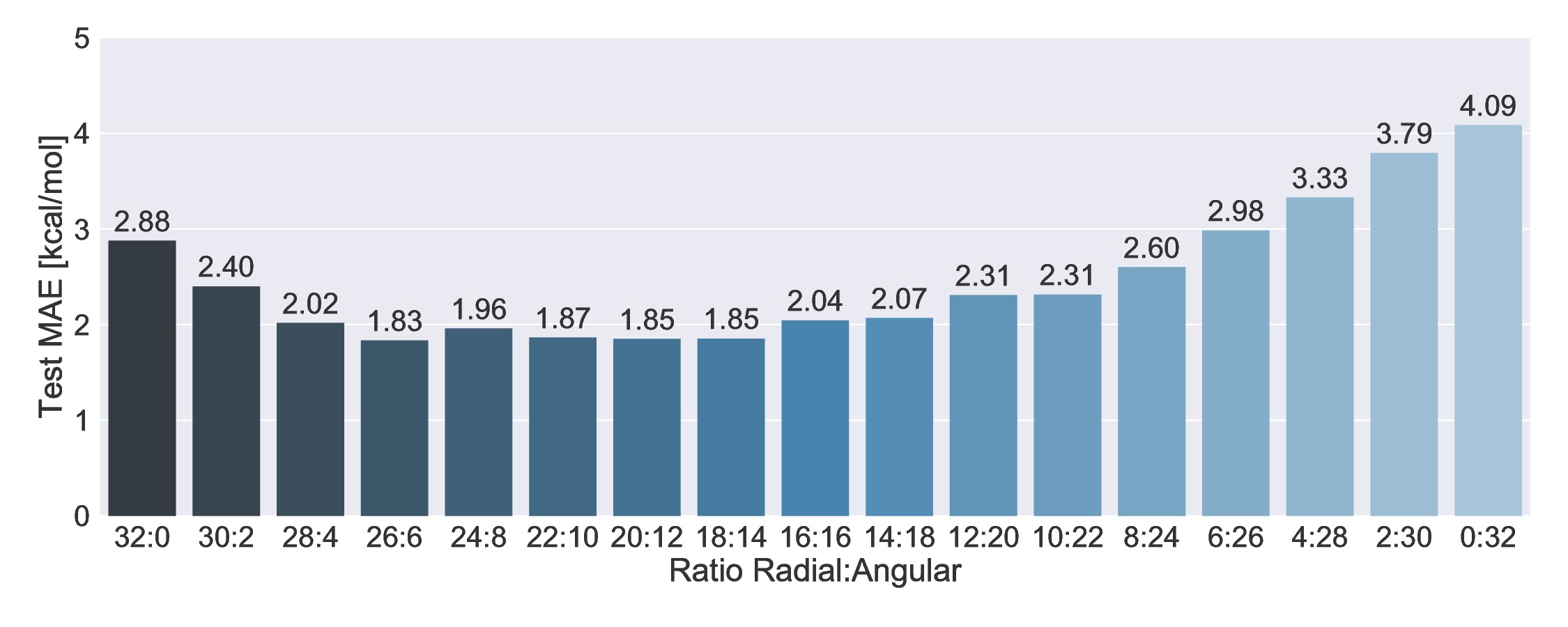}
\caption{Test set MAEs obtained for descriptor vectors using different ratios of radial to angular functions. In all cases, the total number of symmetry functions is kept constant at 32. Since blocks of $\lambda=\pm1$ are used, the number of angular functions can only vary in steps of two.}
\label{fgr:full}
\end{figure}

Figure~\ref{fgr:full} shows all ratios of radial to angular symmetry functions for the weighted atom-centered symmetry function (wACSF) descriptor omitted in the main text.
As can be seen, the general trend of an improvement in accuracy upon adding a few angular functions and deterioration past a certain ratio still holds.

\subsection*{Genetic Optimization of an Additional Set of wACSF.}

\begin{figure}[htpb]
\centering
  \includegraphics[width=0.5\textwidth]{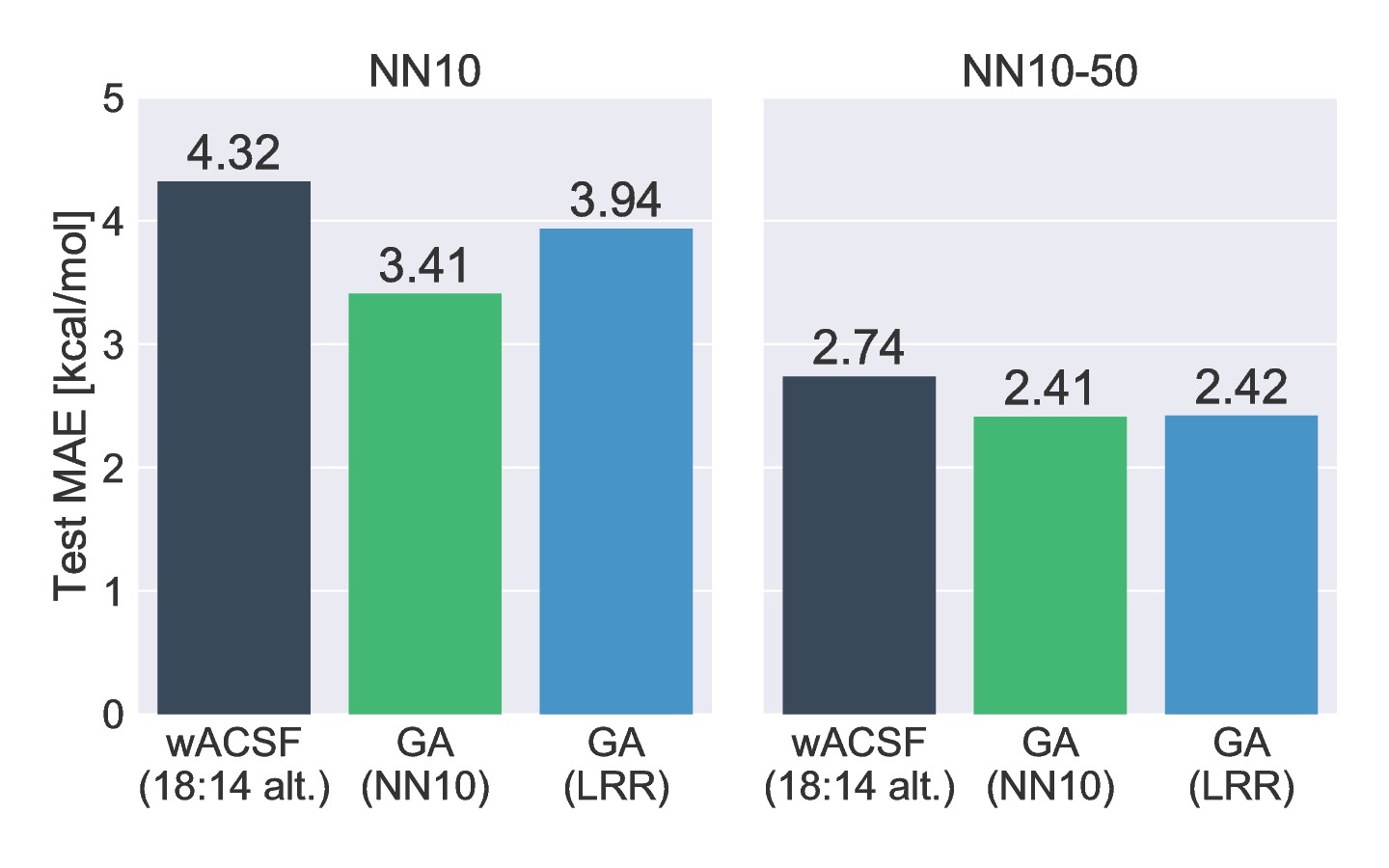}
\caption{Performance of the alternative set of descriptors optimized with the GA using fitness functions based on linear ridge regression (GA-LRR) and HDNNPs (GA-NN10). Shown are the test set MAEs obtained for HDNNP models based on elemental networks using a single layer of 10 nodes (NN10), as well as two layer architectures using 10 and 50 nodes (NN10-50).
The performance of the original descriptor (wACSF, 18:14 alt.) is shown in black. In contrast to the main text, no cross terms depending on $r_jk$ were included in the angular wACSFs of this descriptor.}
\label{fgr:GAalt}
\end{figure}

In addition to the 22:10 radial to angular wACSF vector optimized with a genetic algorithm in the main text, an 18:14 descriptor was optimized. This descriptor is special insofar, as the cross terms depending on $r_jk$ in the angular functions (see Equation~ in the main manuscript) were omitted. The performance of the optimized and unoptimized descriptor vectors is compared in Figure~\ref{fgr:GAalt}. Shown are the results obtained by using a linear ridge regression (LRR) and a high-dimensional neural network potential (HDNNP) based fitness function. In the latter case, single layer elemental networks of 10 nodes were used. Two different types of models were created using these descriptors: a small HDNNP using the same architecture as the HDNNP fitness function (NN10), as well as a larger model employing elemental NNs of two layers with 10 and 50 nodes respectively (NN10-50).
A similar trend as for the 22:10 descriptor is observed. For small models (NN10), the GA leads to a significant improvement of the prediction accuracy, if the HDNNP based fitness score is used as a guidance. This gain diminishes substantially for larger HDNNPs (NN10-50), as the additional capacity of these models can be used to internally transform the presented descriptor vectors to more favorable representations. The results obtained with the LRR based fitness measure show a different behavior than observed in the main text. While no improvement is obtained for the small model, in case of the large model the gain is similar as for the HDNNP based alternative. Based on this dependence on the descriptor to be optimized, we conclude that although it can work in some cases, the LRR fitness function should be used with care.

\subsection*{Comparison of the 22:10 wACSFs pre and post GA optimization.}

\begin{figure}[htpb]
\centering
  \includegraphics[width=\columnwidth]{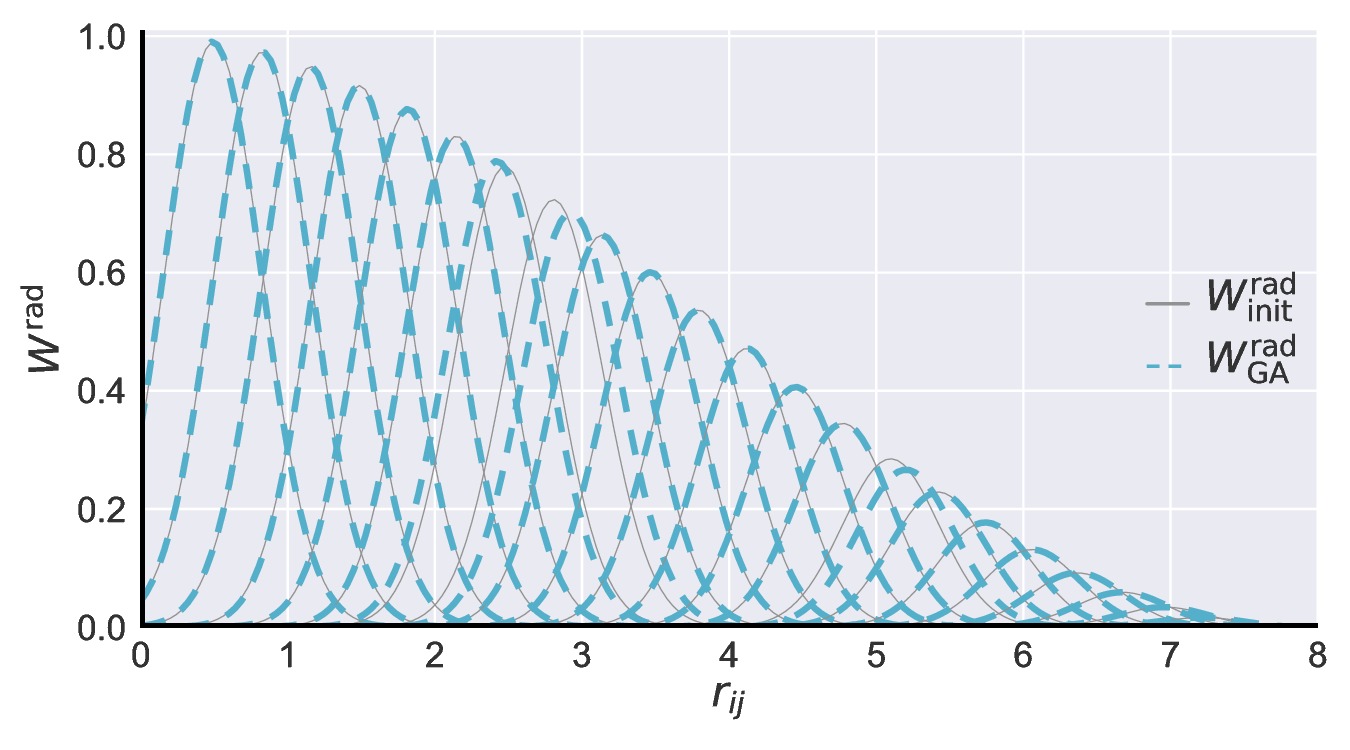}
\caption{}
\label{fgr:rH}
\end{figure}

\begin{figure}[htpb]
\centering
  \includegraphics[width=\textwidth]{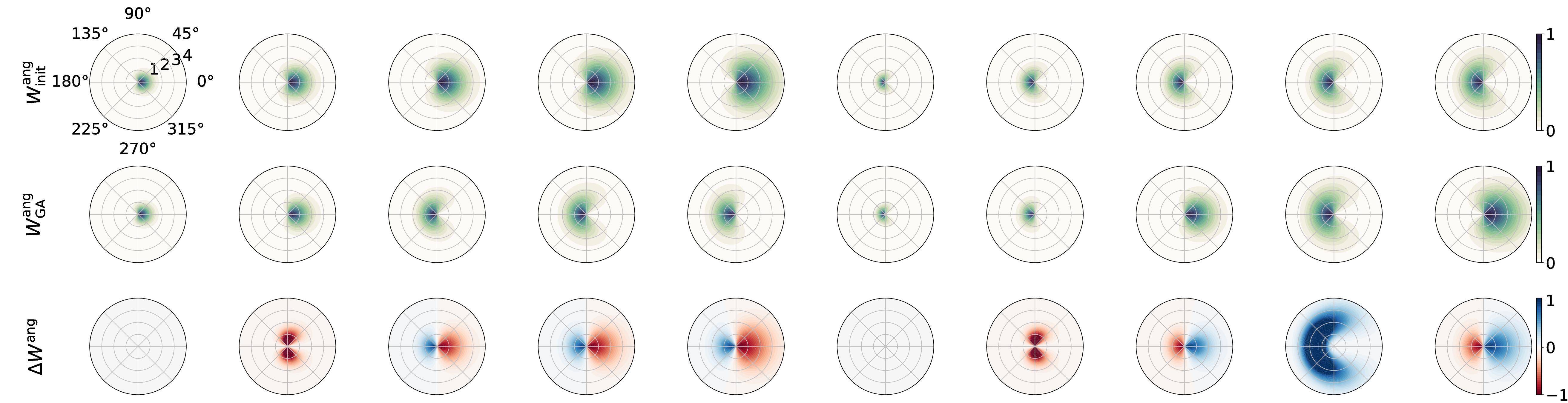}
\caption{}
\label{fgr:aH}
\end{figure}

\begin{figure}[htpb]
\centering
  \includegraphics[width=\columnwidth]{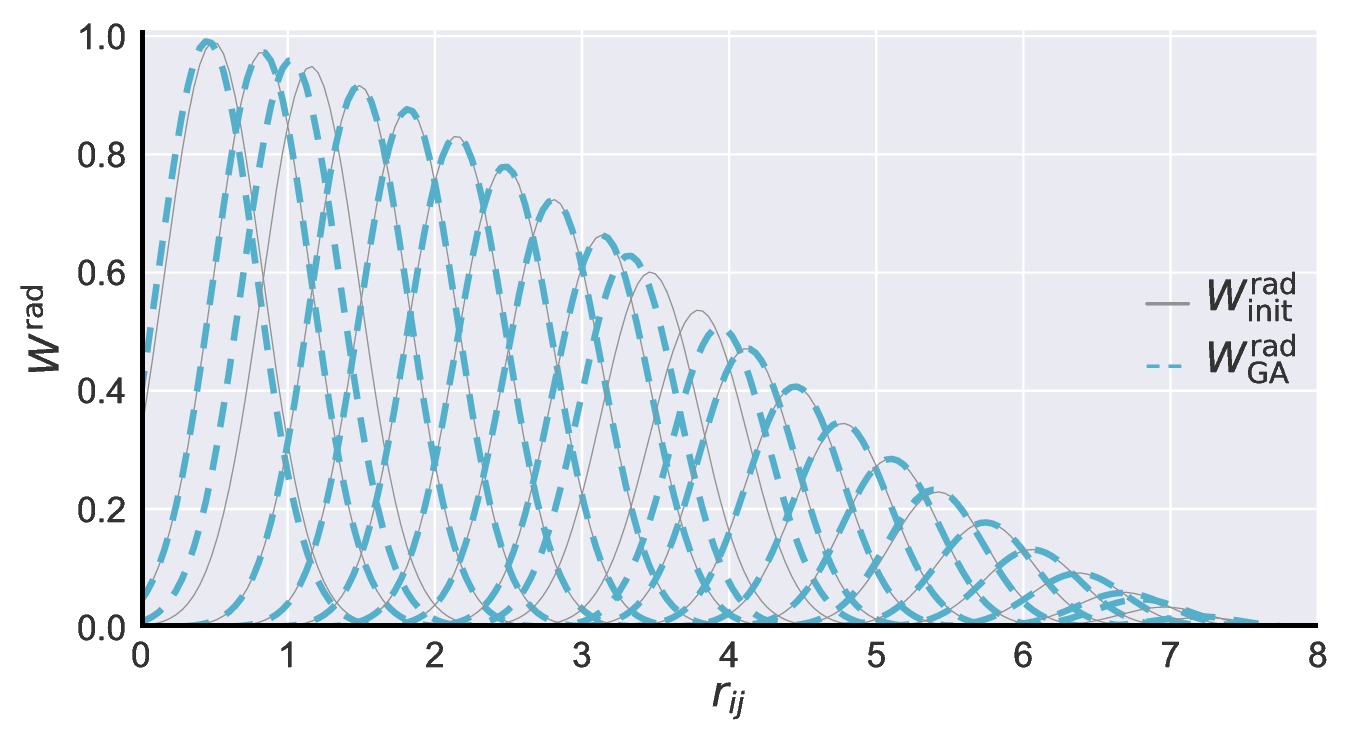}
\caption{}
\label{fgr:rN}
\end{figure}

\begin{figure}[htpb]
\centering
  \includegraphics[width=\textwidth]{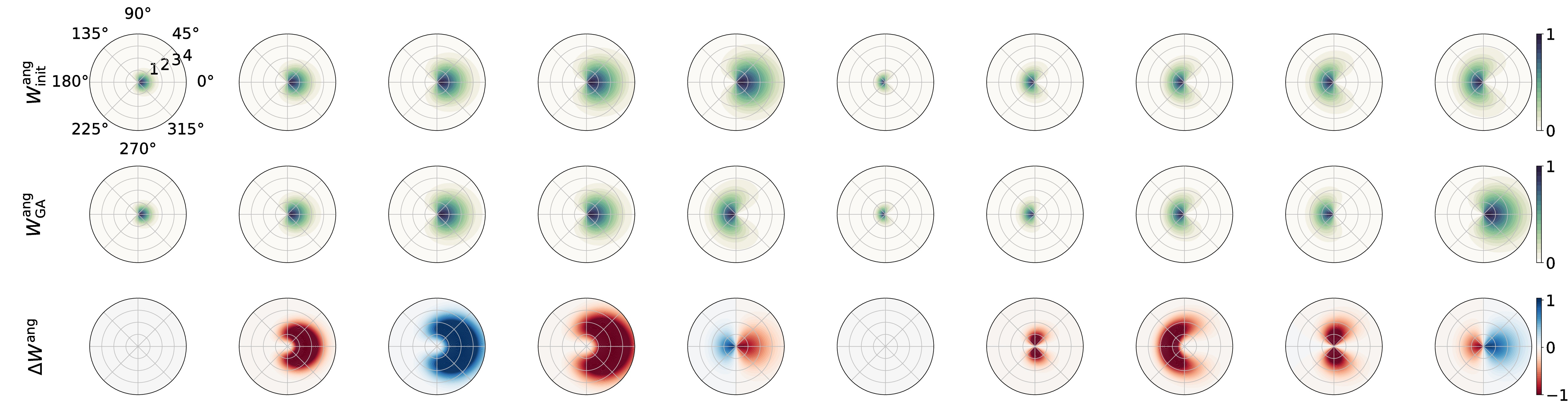}
\caption{}
\label{fgr:aN}
\end{figure}

\begin{figure}[htpb]
\centering
  \includegraphics[width=\columnwidth]{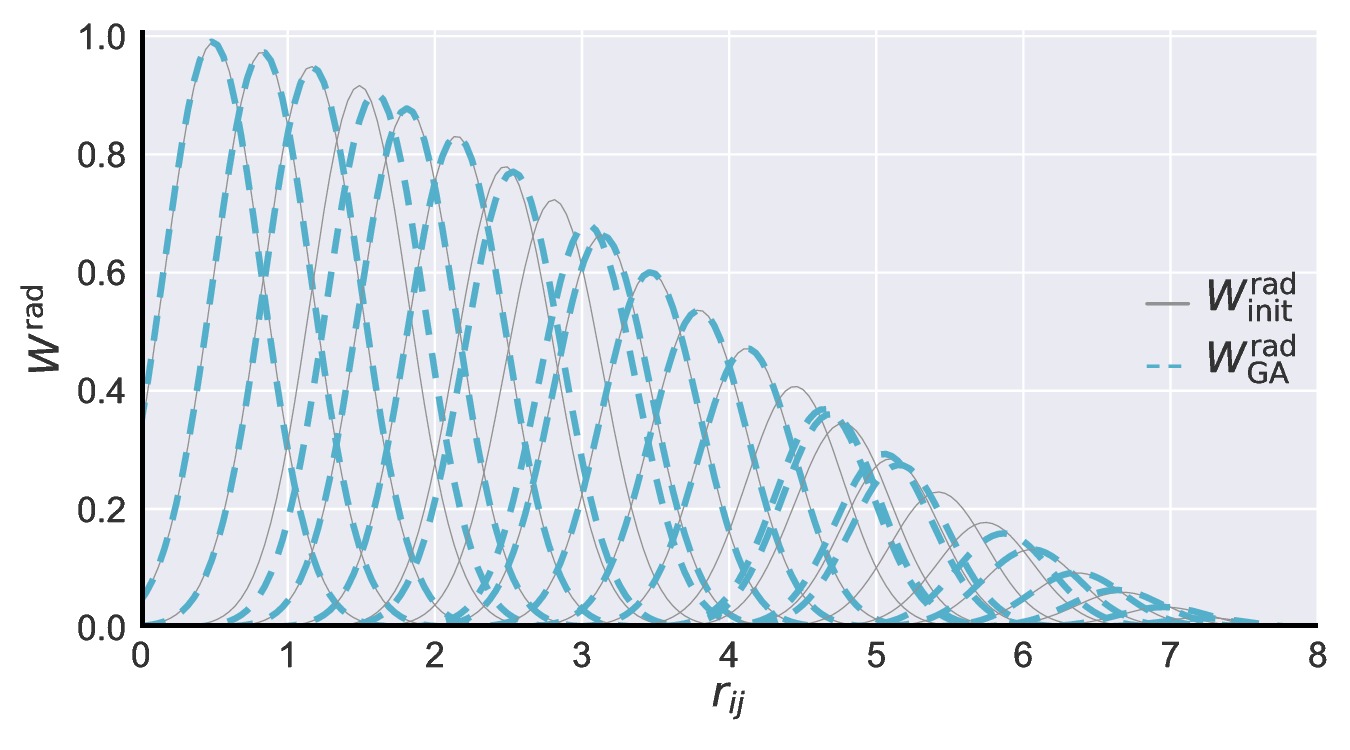}
\caption{}
\label{fgr:rO}
\end{figure}

\begin{figure}[htpb]
\centering
  \includegraphics[width=\textwidth]{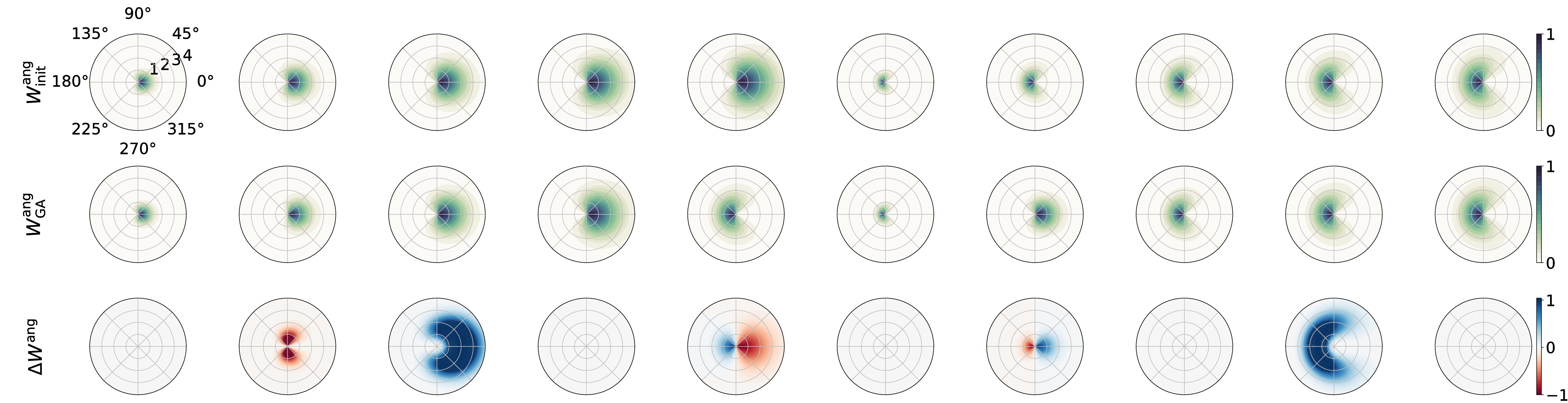}
\caption{}
\label{fgr:aO}
\end{figure}

\begin{figure}[htpb]
\centering
  \includegraphics[width=\columnwidth]{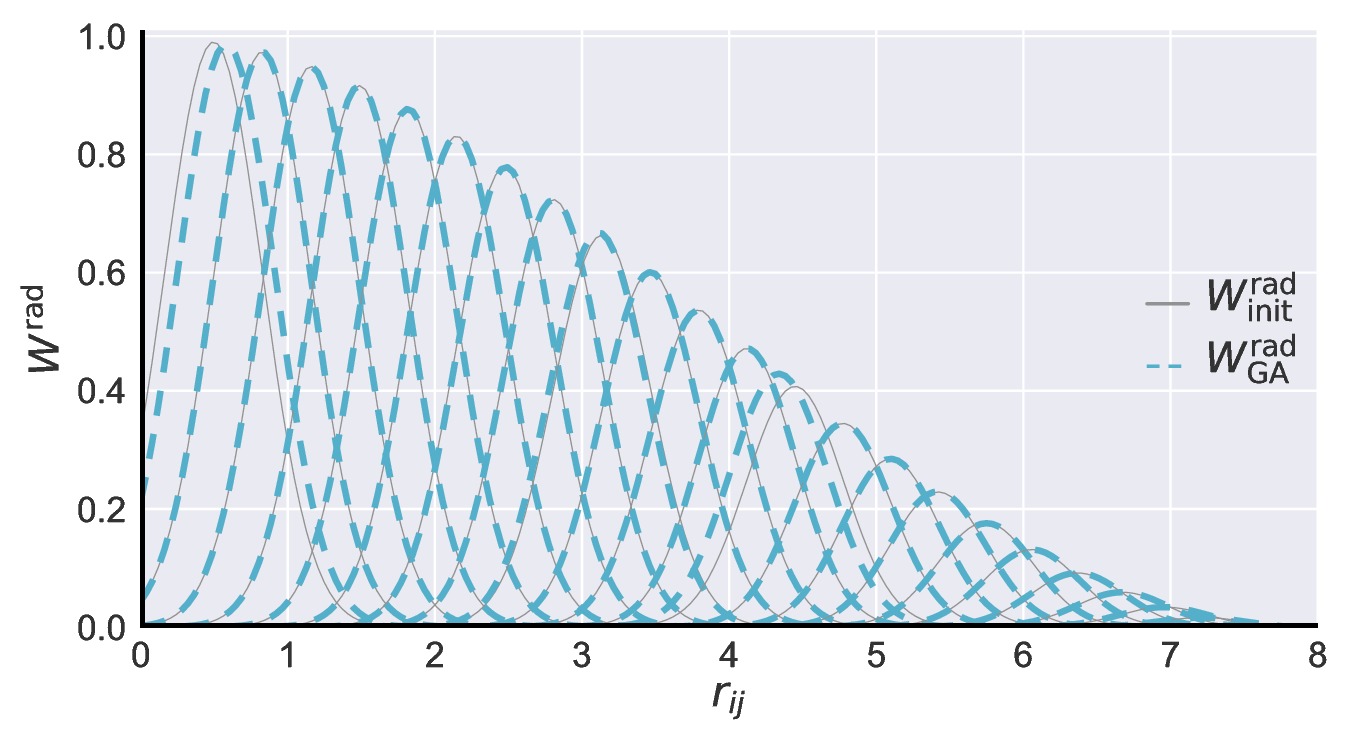}
\caption{}
\label{fgr:rF}
\end{figure}

\begin{figure}[htpb]
\centering
  \includegraphics[width=\textwidth]{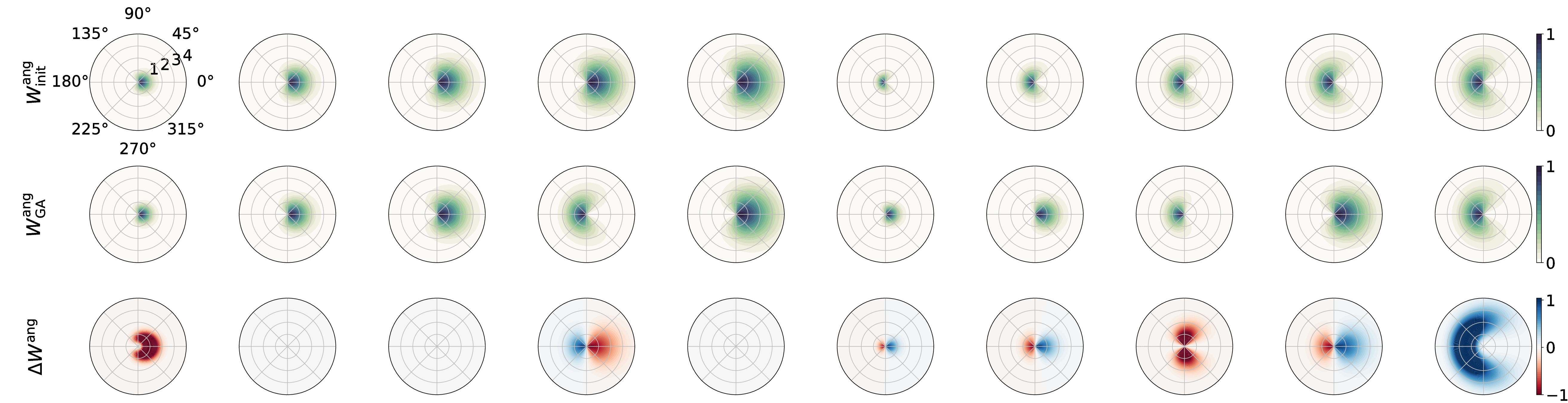}
\caption{}
\label{fgr:aF}
\end{figure}

Here, we present the changes observed in the wACSF 22:10 descriptor during the optimization with the GA for the elements H, N, O and F not shown in the main manuscript. 
The differences in the wACSFs associated with H are given in Figures~\ref{fgr:rH} and \ref{fgr:aH} for the radial and angular functions, respectively.
Figures~\ref{fgr:rN} and \ref{fgr:aN} treat the environments of N atoms, while the descriptors of O are compared in Figures~\ref{fgr:rO} and \ref{fgr:aO}.
Finally, the wACSFs for F are analyzed in Figures~\ref{fgr:rF} and \ref{fgr:aF}.





\end{document}